\documentclass[preprint]{elsarticle}
\biboptions{sort&compress}

\usepackage{graphicx,psfrag,xcolor}
\usepackage{amsmath,amssymb}
\usepackage[list=true, font=normal,
labelformat=brace, position=top]{subcaption}
\usepackage{algorithm,algorithmic}
\usepackage{lineno,hyperref}
\usepackage{array}
\newcolumntype{P}[1]{>{\centering\arraybackslash}p{#1}}

\newcommand{\C}{\mathbb{C}}

\newcommand{\e}{^{(e)}}
\newcommand{\ep}{^{'(e)}}
\newcommand{\es}{^{*(e)}}
\newcommand{\eT}{^{(e)T}}

\newcommand{\Ass}{\overset{m}{\underset{e=1}{\boldsymbol{\mathrm A}}}}

\newcommand{\sty}[1]{\mbox{\boldmath $#1$}}

\newcommand{\ff}{\sty{ f}}
\newcommand{\fp}{\sty{ p}}
\newcommand{\fq}{\sty{ q}}
\newcommand{\fs}{\sty{ s}}

\newcommand{\fu}{\sty{ u}}
\newcommand{\fx}{\sty{ x}}
\newcommand{\fy}{\sty{ y}}
\newcommand{\fz}{\sty{ z}}

\newcommand{\fB}{\sty{ B}}
\newcommand{\fE}{\sty{ E}}
\newcommand{\fK}{\sty{ K}}
\newcommand{\fI}{\sty{ I}}
\newcommand{\fS}{\sty{ S}}

\newcommand{\fsig}{\mbox{\boldmath $\sigma$}}
\newcommand{\feps}{\mbox{\boldmath $\varepsilon $}}
\newcommand{\feta}{\mbox{\boldmath $\eta$}}

\newcommand{\tr}[1]{{\rm tr  }( #1 )}

\begin{document}

\begin{frontmatter}
\title{Efficient Data Structures for Model-free Data-Driven Computational Mechanics}
\author[ifam]{Robert Eggersmann\corref{cor1}}
\ead{robert.eggersmann@ifam.rwth-aachen.de}
\author[icme]{Laurent Stainier}
\ead{laurent.stainier@ex-nantes.fr}
\author[aero,hcm]{Michael Ortiz}
\ead{ortiz@aero-caltech.edu}
\author[ifam]{Stefanie Reese}
\ead{reese@ifam.rwth-aachen.de}
\cortext[cor1]{Corresponding author}
\address[ifam]{Institute of Applied Mechanics, RWTH Aachen University, Mies-van-der-Rohe-Str.1, D-52074 Aachen, Germany.}
\address[icme]{Institute of Civil and Mechanical Engineering, \'{E}cole Centrale de Nantes, 1 Rue de la No\"e, F-44321 Nantes, France.}
\address[aero]{Division of Engineering and Applied Science, California Institute of Technology, Pasadena, CA 91125, USA.}
\address[hcm]{Hausdorff Center for Mathematics, Rheinische Friedrich-Wilhelms-Universit\"at Bonn, Endenicher Allee 62, D-53115 Bonn, Germany.}

\begin{abstract}
The data-driven computing paradigm initially introduced by Kirchdoerfer \& Ortiz \citep{Kirchdoerfer:2016} enables finite element computations in solid mechanics to be performed directly from material data sets, without an explicit material model.
From a computational effort point of view, the most challenging task is the projection of admissible states at material points onto their closest states in the material data set.
In this study, we compare and develop several possible data structures for solving the nearest-neighbor problem.
We show that approximate nearest-neighbor (ANN) algorithms can accelerate material data searches by several orders of magnitude relative to exact searching algorithms.
The approximations are suggested by---and adapted to---the structure of the data-driven iterative solver and result in no significant loss of solution accuracy.
We assess the performance of the ANN algorithm with respect to material data set size with the aid of a 3D elasticity test case.
We show that computations on a single processor with up to one billion material data points are feasible within a few seconds execution time with a speed up of more than $10^6$ with respect to exact $k$-d trees.
\end{abstract}

\end{frontmatter}

\section{Introduction}\label{sec:intro}

The classical paradigm of computational mechanics is to use data from experimental tests to formulate material models by fitting to the data, then use the models in calculations.
The process of modeling is often ill-posed and open-ended, results in loss of information relative to the original material data set and introduces epistemic uncertainty into the calculations.
As material data becomes more plentiful, owing to advances in experimental science, multiscale modeling and other data sources, a new paradigm, model-free Data-Driven (DD) computational mechanics, suggests itself.
The objective of the {DD} paradigm, first introduced in \cite{Kirchdoerfer:2016}, is to make predictions regarding the behavior of physical systems directly from material data, without the intermediate step of material modeling.

{DD} entails a reformulation of the initial boundary value problems in which the solution is sought within a set of admissible states, or \textit{constraint set}, and required to minimize distance to the material data set. 
The constraint set spans all states in phase space, e.~g., stress and strain, that satisfy equilibrium and kinematic relations, as well as boundary conditions.
{DD} has been extended to noisy data sets \cite{Kirchdoerfer:2017}, dynamics \cite{Kirchdoerfer:2018}, inelasticity \cite{Eggersmann:2019}, finite deformations \cite{Nguyen:2018, Platzer:2019, Conti:2020}, fracture \cite{Carrara:2020}, and second-order approximations \cite{Eggersmann:2020}.
Leygue et al. (2018) \cite{Leygue:2018, Dalemat:2019, Leygue:2019, Stainier:2019} proposed a closely-related inverse approach, termed Data-Driven Material Identification (DDMI), that takes strain fields from full-field optical measurements and corresponding loading boundary conditions to infer the corresponding stress field. 
DDMI can be used to generate very large material data sets directly from a small number of full-field microscopy measurements.\\

DD solvers, such as proposed in \cite{Kirchdoerfer:2016}, iteratively minimize the distance between the material data set and the constraint set.
From the standpoint of time efficiency, the central step in the solver is searching for the point in the material data set nearest to a given local state.
Evidently, this search is an instance of the classical nearest-neighbor problem.
The nearest-neighbor problem is also referred to as the post-office problem and can be solved by means of many algorithms.
In general, the problem is defined by finding the closest point in a given set to a query point, where both the elements of the given set and the query point are points in the same space.
A generalization of the problem is the $k$-nearest-neighbor problem, which is also closely related to the fixed-radius nearest-neighbor problem and the all nearest-neighbor problem.

The simplest algorithm for finding the exact solution to the nearest-neighbor problem is the linear search, where the query point is compared to every point in the data set.
This naive approach is prohibitively expensive for very large data sets.
Another possibility is the use of space partitioning algorithms where the branch and bound method \cite{Land:2010} is applied.
An early example is the $k$-d tree developed by Bentley \& Friedman in 1975 \cite{Friedman:1977,Bentley:1975}.
Here, $k$ represents the dimension of the data set, and the search space is recursively partitioned along one dimension in each branching step.
Another early algorithm is the $k$-means tree investigated by Fukunaga \& Narendra in 1975 \cite{Fukunaga:1975}, where $k$ stands for the number of clusters built into each branch.
Other algorithms include the Quad-tree \cite{Finkel:1974}, the R-tree \cite{Guttman:1984}, the metric-tree \cite{Uhlmann:1991, Yianilos:1993,Ciaccia:1997}, the ball-tree \cite{Omohundro:1989}, among many others, with numerous variants thereof.
A deficiency of exact methods is that they cannot guarantee logarithmic search time, as shown in \cite{Uhlmann:1991} or \cite{Sproull:1991}.\\

Conveniently, for many applications it is sufficient to find a good guess close to the true nearest-neighbor. 
These approximate nearest-neighbor (ANN) algorithms have recently attracted much attention since they are of practical importance for many applications in sales or social media. 
The main idea behind ANN algorithms is to exploit the trade-off between efficiency and accuracy.
In an early contribution, Miclet \& Dabouz \cite{Miclet:1983} investigated a hierarchical search structure based on $k$-means with no backtrack. 
Arya \textit{et al.} \cite{Arya:1993,Arya:1998} introduced the $(1+\varepsilon)$-nearest-neighbor criterion.
A $(1+\varepsilon)$-nearest neighbor is a point, whose distance to the query point is less than $(1+\varepsilon)$ times the distance to the true nearest neighbor.
Brin \cite{Brin:1995} proposed a geometric near-neighbor access tree (GNAT) as an extension of $k$-means trees. 

An approach based on randomized $k$-d trees was proposed in \cite{Silpa:2008}, where the search space is partitioned multiple times and no backtracking is performed. 
Muja and Lowe \cite{Muja:2009a} further compared the randomized $k$-d trees with a hierarchical $k$-means tree using a bounded priority queue for backtracking. 
This idea of using priority queues for backtracking, also called the best-bin first (BBF) method,  was initially proposed in \cite{Beis:1997} for $k$-d trees. 
The nodes to be checked during backtracking are stored in a sorted queue that checks the nodes with the smallest bin distances first. 
The length of the priority queue is then limited to be less than a prescribed number. 
Muja and Lowe also launched the FLANN-library \cite{Muja:2009b}, which is well-established today. 
Further investigations on the scalability of the algorithms were presented in \cite{Muja:2014}. 

More recent work has focused on proximity graph-based methods for high-dimensional problems.
Those problems arise, e.~g., in image recognition, computational linguistics, or product recommendation. Graph-based methods seem to be superior in those fields. 
Early graph-based searching methods were based on monotonic search networks introduced by Dearholt \textit{et al.} \cite{Dearholt:1988} and on randomized neighborhood graphs by Arya and Mount in 1993 \cite{Arya:1993}. 
Hajebi et al. \cite{Hajebi:2011} do the nearest-neighbor search by means of a hill-climbing algorithm in a graph where every node is linked with its $k$ nearest-neighbors. 
Dong \textit{et al.} \cite{Dong:2011} proposed an algorithm to construct the $k$-NN graphs efficiently by an iterative procedure. 
Other efficient graph search algorithms are the navigable small world graph \cite{Malkov:2014} and the hierarchical navigable small world (HNSW) graph \cite{Malkov:2018}. 
In these methods, a small world graph is an approximation of a Delaunay graph. 
The HNSW graph has multiple layers. 
The bottom layer includes all points. 
With increasing layer number, the number of points decreases. 
These graph-search algorithms are compared with the diversified proximity graph, which is based on an existing $k$-NN graph, in \cite{Li:2019}. 
In \cite{Fu:2017}, a navigating spreading-out graph is proposed that enables billion-scale data set searches for online sales applications. 
The same authors introduce the satellite system graph (SSG) in \cite{Fu:2019}. 
This graph regulates its sparsity by a minimal angle of the vectors to its neighbors.
Recently, \textit{Groh et al.} \cite{Groh:2019} proposed a graph-based GPU nearest-neighbor algorithm that betters the currently most efficient algorithms by more than a magnitude in search time.

In general, choosing the best search algorithm depends on many parameters. 
Data-related properties such as the dimension, number of examples, correlations, and density distributions need to be carefully considered. 
In addition, the objectives of the user must to be taken into consideration. 
Relevant considerations are query time, accuracy, query workload, building time and memory usage.\\

Our present work focuses on the suitability and performance of different nearest-neighbor search algorithms in the context of Data-Driven computing with noise-free material data sets of up to a billion points. 
We investigate tree-based and graph-based methods. 
The aim of the work is to ascertain how the specific features of DD iterative solvers can be best exploited to accelerate searches. An examination of the iterative solvers suggests two lines inquiry:

\begin{enumerate}
\item We observe that it is sufficient to use rough guesses for the nearest-neighbor search in the initial DD iterations during which the accuracy of ANN algorithms can be set low.
With decreasing distance to the solution, the accuracy of the searches needs to be steadily increased.
\item We additionally observe that the query points move relatively little when approaching convergence. 
In the context of graph search algorithms, this effect represents an increasing amount of knowledge that can be used to choose better starting points for navigating through the search graph.
\end{enumerate}

We show that, by exploiting these features, approximate nearest-neighbor (ANN) algorithms can accelerate material data searches by several orders of magnitude relative to exact searching algorithms. 
We emphasize that the approximations are suggested by---and adapted to---the structure of the data-driven iterative solver and result in no significant loss of solution accuracy. 
We additionally assess the performance of ANN algorithms with increasing material data set size with the aid of a 3D elasticity test case. 
We show that computations on a single processor with up to one billion material data points are feasible within a few seconds execution time with a speed up of more than $10^6$ with respect to exact $k$-d trees.

The paper is structured as follows.
In Section 2, we recapitulate the {DD} paradigm and its iterative solution procedure. 
The efficiency of a 3D elastic solid example with varying data set sizes using a $k$-d tree is investigated in Section 3. 
In addition, we investigate the characteristics of the nearest-neighbor search that appear in the data-driven calculation. 
In Section 4, we study different ANN algorithms for accelerating computations and assess their performance on material data sets of up to one million points, with specific focus on the trade-off between accuracy and speed. 
A comparison of the different ANN algorithms with the most efficient parameters follows. 
These comparisons are based on data sets of up to 100 million points. 
Finally, we show that computations with billion-point data sets are possible within seconds.
Section 5 closes with final conclusions and outlook.

\section{The iterative solver of the data-driven problem}
\label{sec:datadriven}

The data-driven formulation of the discretized initial boundary value problem of
elasticity, as proposed in [1, 5], can be stated as follows.
A system undergoes displacements $\fu = \{\fu_i\}^n_{i=1}$, with $\fu_i \in \mathbb{R}^{n_i}$ being the
displacement vector at node $i$ with dimension $n_i$ at all nodes $i = 1, ..., n$, under the action of applied forces $\ff = \{\ff_i\}_{i=1}^n$.
The vector $\ff_i \in \mathbb{R}^{n_i}$ denotes the nodal force vector. \\
%
The following minimization then defines the data-driven problem
\begin{equation}\label{eq:ddProblem}
\min_{\fy \in C}\min_{\fz \in D} d^2(\fy,\fz),
\end{equation}
i.~e., the objective is to find the state $\fy$ in the constraint set $C$, which is closest to the state $\fz$ in the data set $D$, which is again closest to the constraint set.
Here, the constraint set $C$, as well as the data set $D$, are subsets of the global phase space $Z$.
The constraint set contains all admissible states fulfilling equilibrium and kinematic relations.
Experimental measurements or computations of micro-structures sample the data set.
The associated squared distance $d^2(\fy,\fz)$ in phase space is defined by
\begin{equation}
d^2(\fz,\fy)=\sum_{e=1}^m \frac{1}{2} w\e d^2_e(\fz\e,\fy\e),
\end{equation}
where $w_e$ are volumes associated with the integration points $e = 1,\dots, m$ with $\fy\e=(\feps\e,\fsig\e)$, $\fz\e=(\feps\ep,\fsig\ep)$ being states in the local phase space $Z\e\in\mathbb R^{2M}$.
The corresponding local distance is then defined by
\begin{equation}
d^2_e(\fy\e,\fz\e)=\mathbb{C}\e(\feps\e-\feps\ep)\cdot(\feps\e-\feps\ep)+\mathbb{C}^{(e)-1}(\fsig\e-\fsig\ep)\cdot(\fsig\e-\fsig\ep).\label{eq:distmetric}
\end{equation}
Here, the metric $\mathbb{C}\e$ is a symmetric positive definite matrix bringing the absolute values of stress and strain to an equal scale.\\
%

The solution scheme proposed in \cite{Kirchdoerfer:2016} then iteratively projects a state $\fz_{i} \in D$ with $\fz\e_i=(\feps\es_i,\fsig\es_i)$ to the closest point in the constraint set $\fy^{i+1} \in C$, where the index $i$ indicates the current iteration.
\begin{equation}
\fy_{i+1}=P_C(\fz_{i}).
\end{equation}
\begin{figure}[htbp]
\begin{subfigure}{0.5\textwidth}
\centering
\psfrag{sig}{$\sigma$}
\psfrag{eps}{$\varepsilon$}
\psfrag{C}{$C$}
\psfrag{zi}{$\fz_i$}
\psfrag{yi}{$\fy_i$}
\psfrag{Pc}{$P_C$}
\psfrag{A}{$A$}
\psfrag{F}{$F$}
\includegraphics[width=\textwidth]{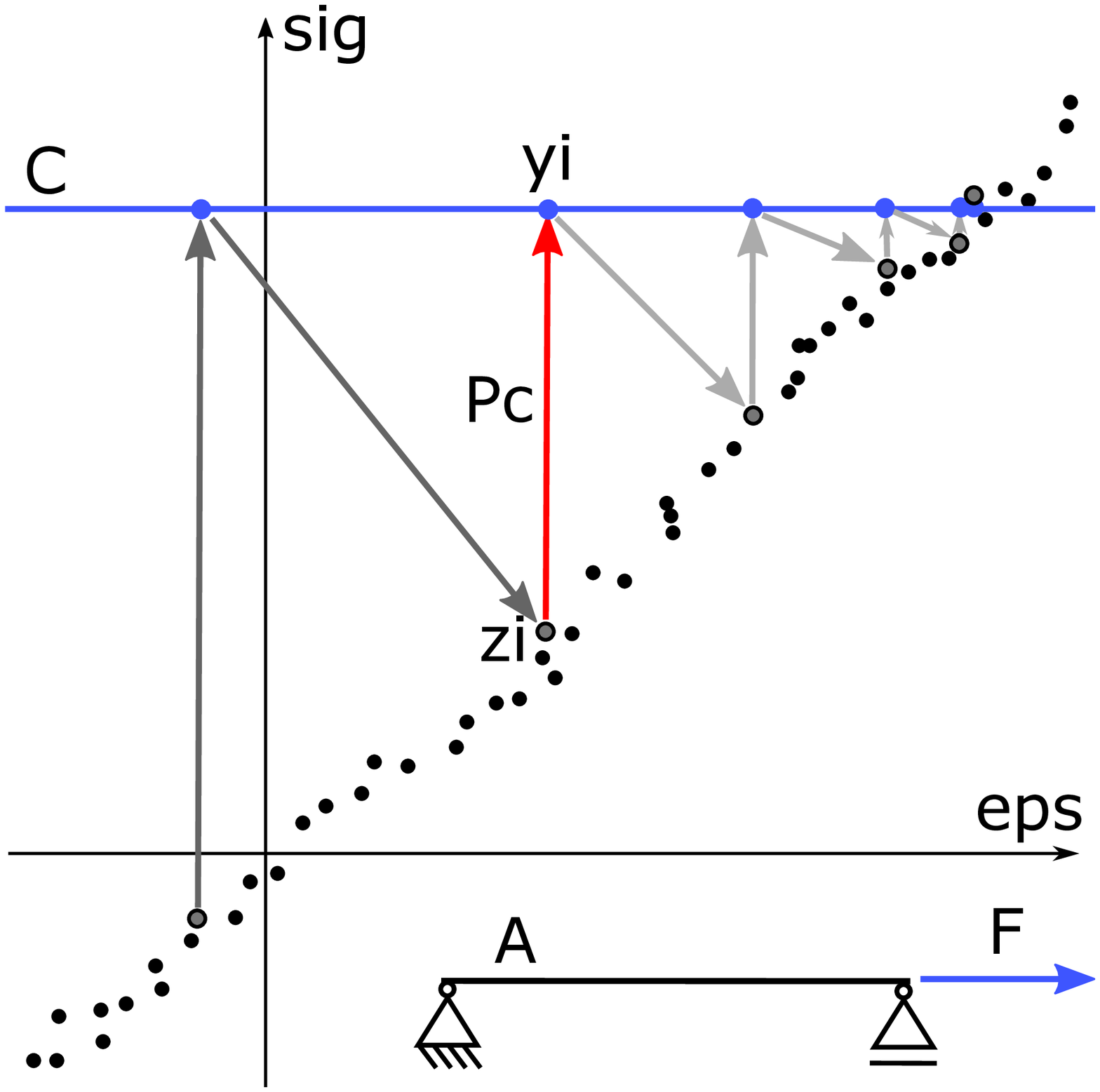}
\label{fig:projectionC}
\end{subfigure}
\begin{subfigure}{0.5\textwidth}
\centering
\psfrag{sig}{$\sigma$}
\psfrag{eps}{$\varepsilon$}
\psfrag{C}{$C$}
\psfrag{zi}{$\fz_{i+1}$}
\psfrag{yi}{$\fy_i$}
\psfrag{Pd}{$P_D$}
\includegraphics[width=\textwidth]{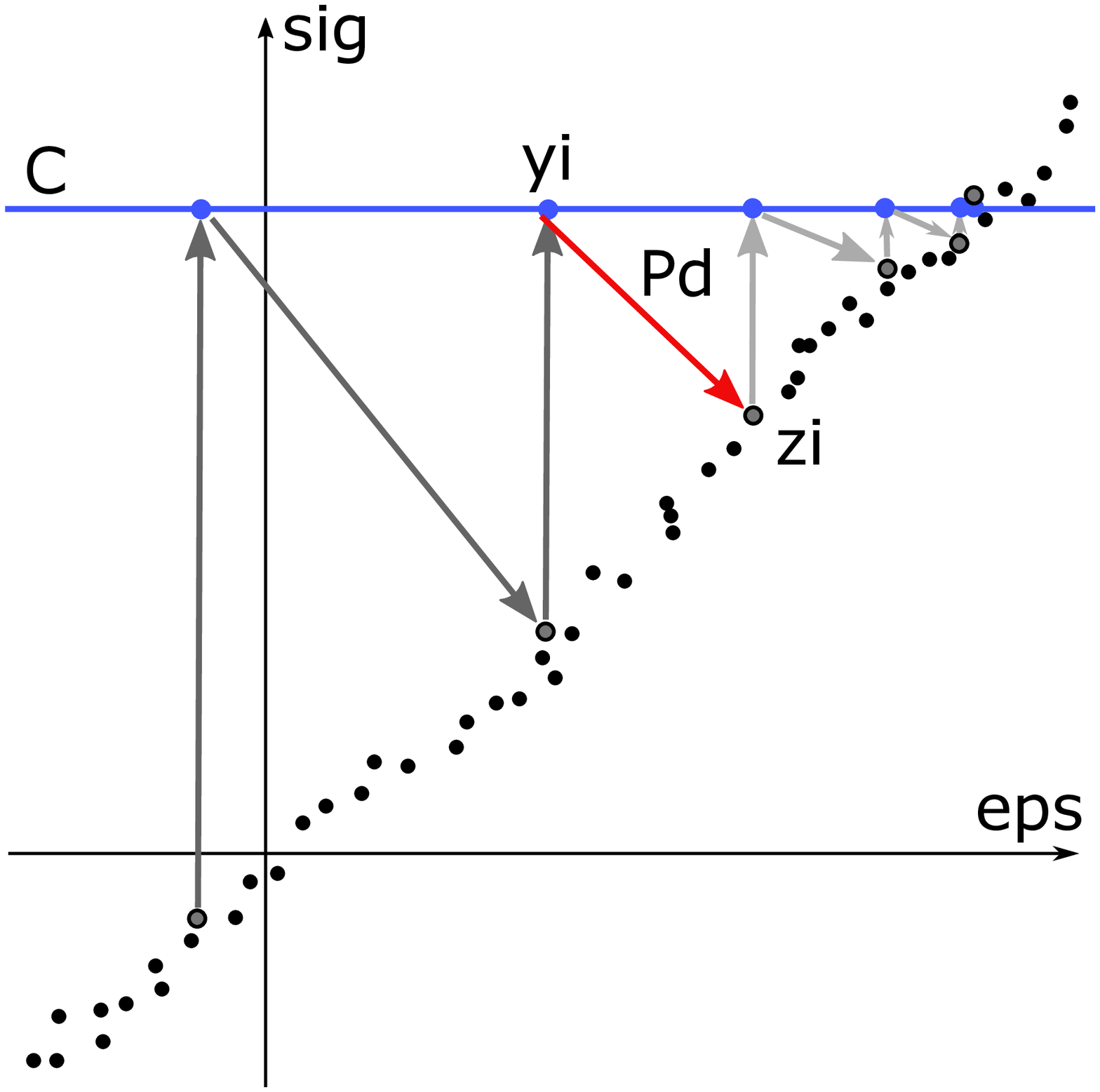}
\label{fig:projectionD}
\end{subfigure}
\caption{Illustration of the data-driven iterative solver for a single truss problem with load $F$ and cross section area $A$. Data set (black points) and constraint set (blue line) are given in the $\varepsilon$-$\sigma$-phase space. Left side: Projection $P_C$ of state $\fz_i$ in the constraint set. Right side: Projection $P_D$ of state $\fy_i$ in the data set $D$.}
\label{fig:projections}
\end{figure}
The projection into the constraint set $C$ is performed by solving the two equation systems
\begin{subequations}
\begin{equation}\label{eq:KuE}
\underbrace{\Ass\{w\e\fB\eT\mathbb{C}\e\fB\e\}}_{\fK}\fu=\underbrace{\Ass\{w\e\fB\eT\mathbb{C}\e\feps\es\}}_{\fE},
\end{equation}
\begin{equation}\label{eq:KeS}
\underbrace{\Ass\{w\e\fB\eT\mathbb{C}\e\fB\e\}}_{\fK}\feta=\underbrace{\ff-\Ass\{w\e\fB\eT\fsig\es\}}_{\fS},
\end{equation}
\end{subequations}
and computing the new values for stress and strain by
\begin{subequations}
\begin{equation}
\feps\e=\fB\e\fu\e, \hspace{1cm} \forall e=1,...,m,
\end{equation}
\begin{equation}
\fsig\e=\fsig\es+\mathbb{C}\e\fB\e\feta\e, \hspace{1cm} \forall e=1,...,m.
\end{equation}
\end{subequations}
where $\fB\e$ denotes the classical strain-displacement operator. \\
Regarding the computational implementation it is reasonable to factorize the stiffness matrix $\fK$ directly after the first assembly.
Then, for geometric linear and constant $\mathbb C\e$, only $\fE$ and $\fS$ have to be reassembled in every iteration.\\

The second part of a single iteration is to find the closest state in the data set to the previously calculated state in the constraint set.
This second projection
\begin{equation}
	\fz_{i+1}=P_D(\fy_{i+1})
\end{equation}
then specifies to minimize the local distances $d_e$ for given states $\fy\e_{i+1}$ at all integration points $e$.
In other words, a nearest-neighbor problem has to be solved for each integration point, where we aim to find the state in the data set $\fz\e_{i+1}\in D_e=\{\feps\ep_j,\fsig\ep_j\}_{j=1}^N$ closest to $\fy\e_{i+1}$ regarding the metric \eqref{eq:distmetric}.
The data set's size is specified by the cardinality $N$ and dimension $\dim(Z\e)$.
The latter is the sum of the $M$ stress and $M$ strain components in Voigt notation.\\
One iteration of the {DD} solver can be finally expressed by
\begin{equation}
\fz_{i+1}=P_D\big(P_C(\fz_i)\big).
\end{equation}
Starting the solver, the integration points are randomly associated with points from the data set.
The final result of the problem is then chosen to be the state in the constraint set obtained when the global distance does not decrease any more.
Equivalently, if the nearest-neighbor projection is exact, no change in the integration point's association with the data points will finally be observed.\\
%
%
Here, it should be remarked that the iterative solver with the minimum distance formulation does not find the best solution to the data-driven problem described in Eq.~\ref{eq:ddProblem} in general.
This is because the solver can stop at a local minimum if distances between points are too large.
However, since this effect usually appears only quite close to the true solution, the obtained solution is usually a good approximation.
It is possible to reduce this effect by using e.~g., the maximum-entropy formulation \cite{Kirchdoerfer:2017} or a local second-order approximation described in \cite{Eggersmann:2020} or a combination of both.

\section{Example: 3D elastic solid with exact nearest-neighbor projections}
\label{sec:bvp}
\begin{figure}[htbp]
\centering
 \psfrag{a}{$\theta=2^\circ$}
\includegraphics[width=0.4\textwidth]{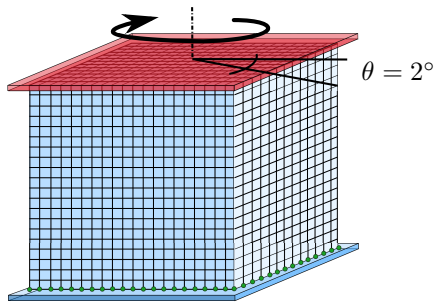}
\caption{Elastic solid cube with discretization and boundary conditions. Nodes at the lower blue surface are fixed. Nodes at the upper red surface are rotated around the centered z-axis and fixed in z-direction.}
\label{fig:cube}
\end{figure}

In this work, all investigations are based on the problem of a three-dimensional elastic solid described in the following.
Nevertheless, the phenomena, which will be discussed here, also occur with problems of lower-dimensional data sets like trusses or continua in plane strain or plane stress conditions.
Furthermore, the following search algorithms can be applied on problems in dynamics \cite{Kirchdoerfer:2017} and inelasticity \cite{Eggersmann:2019} in the same way.\\
The boundary value problem considered is a cube of side length $10$ mm discretized by $20 \times 20 \times 20$ elements.
The degrees of freedom at the bottom are fixed in all directions, whereas the cube's upper surface is rotated by $2^\circ$, as depicted in Fig.~\ref{fig:cube}.
In total, 64000 integration points are evaluated in each iteration.\\
The reference material behavior is given by an isotropic but nonlinear material description of the form:
\begin{equation}\label{eq:material}
\fsig=E\big(\feps+\alpha\feps^3+0.5\,(\tr{\feps}+\alpha\,\tr{\feps}^3)\fI\big),
\end{equation}
where $E=1000\,\text{MPa}$ and $\alpha=500$ are material parameters. \\
Data sets $D\e=\{(\feps_i,\fsig_i)\}_{i=1}^N$ of dimension twelve and N artificial measurements are randomly created on a range $[-0.025;\,0.025]$ for all strain components in Voigt notation.
The corresponding stresses are computed by the reference material model (Eq.~ \ref{eq:material}). \\
Here, the constant matrix $\mathbb C\e$ is determined by using a principal component analysis applied to the data.
Therefore, the first $6$ principal component vectors are determined and written in a $(12 \times 6)$-matrix $\boldsymbol A$.
Then $\mathbb C\e=\text{sym}(\boldsymbol A_{\varepsilon}^{-1}\boldsymbol A_{\sigma})$, where $\boldsymbol A_{\varepsilon}$ is the upper and $\boldsymbol A_{\sigma}$ the lower $(6 \times 6)$-matrix of $\boldsymbol A$.\\
%

\subsection{Performance studies using an exact 12-d tree.}\label{sec:exactkd}
\begin{figure}[!bp]
\begin{subfigure}{0.5\textwidth}
\centering
\includegraphics[width=\textwidth]{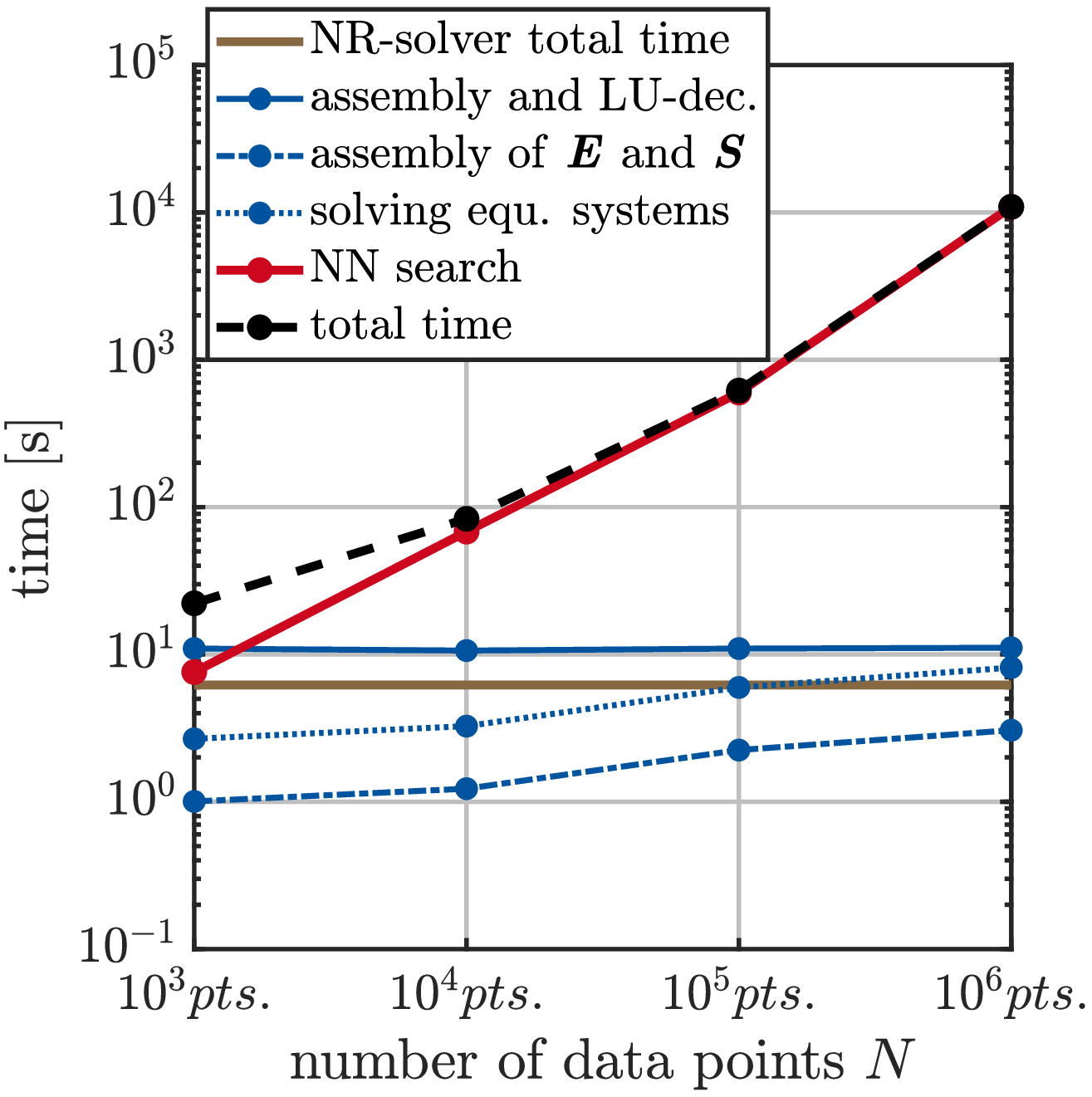}
\subcaption{}
\label{fig:timePoints}
\end{subfigure}
\begin{subfigure}{0.5\textwidth}
\centering
\includegraphics[width=\textwidth]{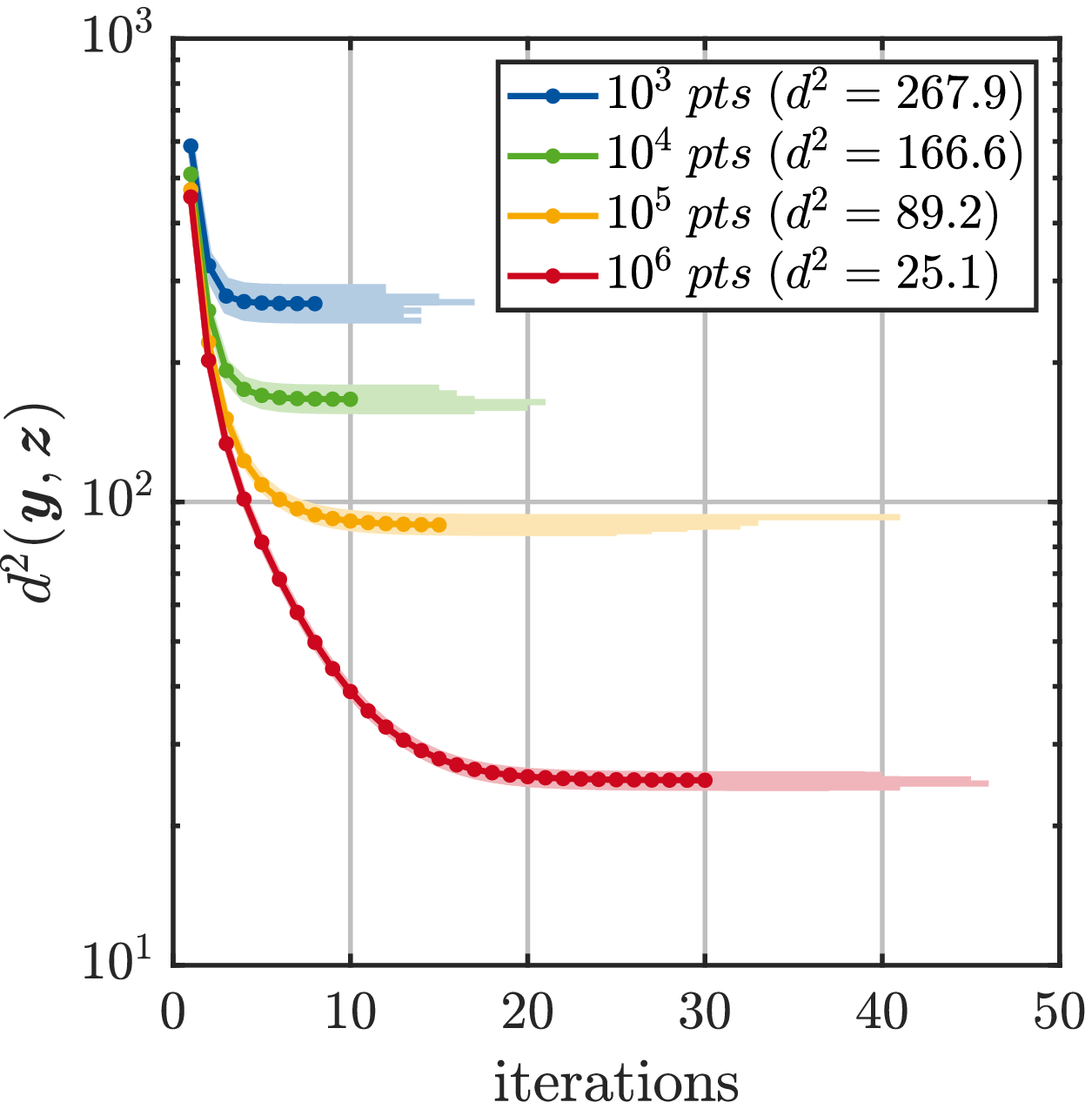}
\subcaption{}
\label{fig:distIterPoints}
\end{subfigure}
\caption{Elastic solid example: 12-d tree used as data structure for exact nearest-neighbor search. a) Comparison of computational times for different data set refinements. Times are averaged over the 20 samples. b) Global squared distance over solver iterations for different data set refinements. Averaged distances in bold, all results in light colors. Averaged final distances after 8, 10, 15 and 30 iterations stated in brackets.}
\label{fig:kD_points}
\end{figure}

\begin{figure}[htbp]
\begin{subfigure}{0.5\textwidth}
\centering
\includegraphics[width=\textwidth]{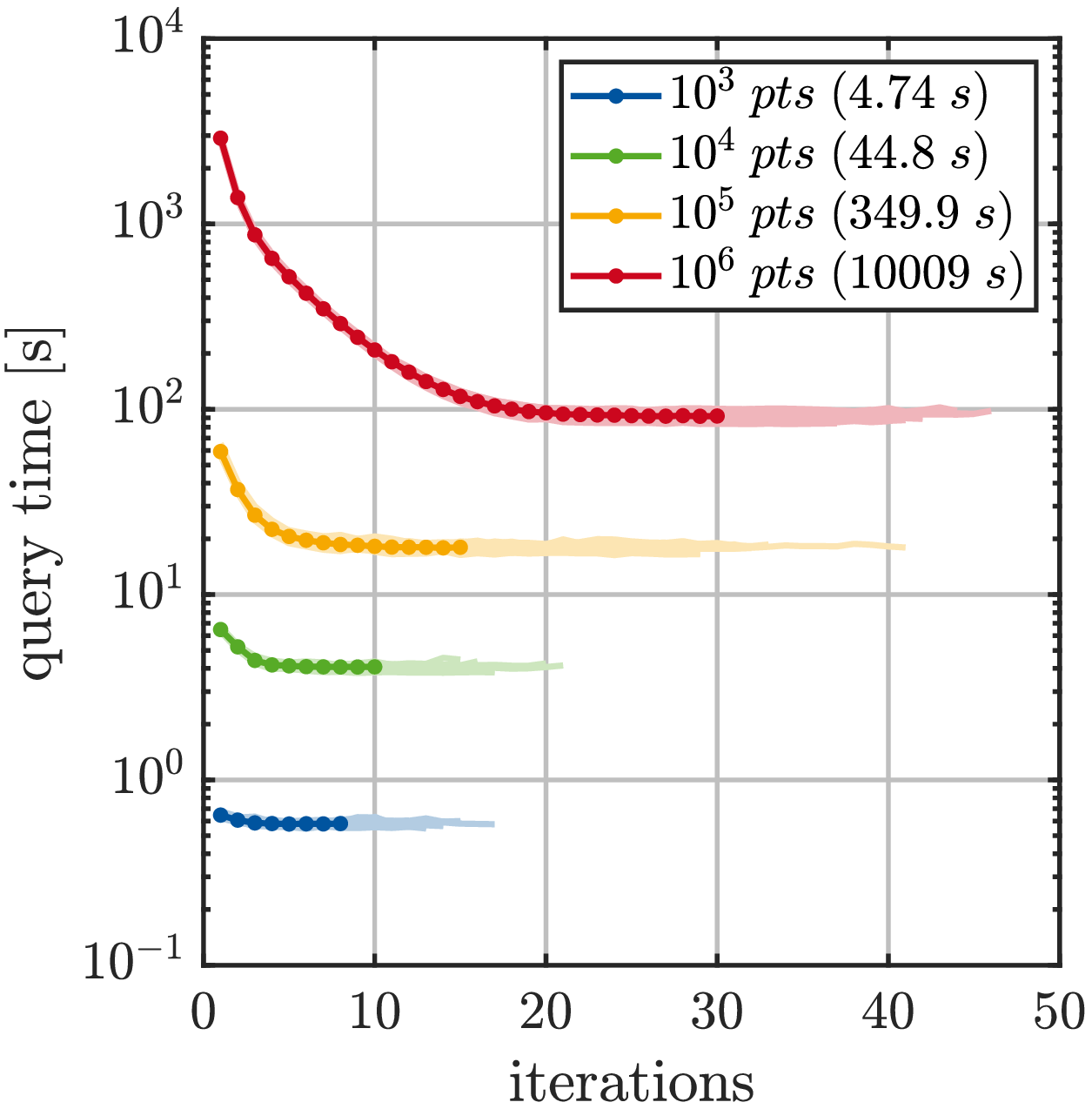}
\subcaption{}
\label{fig:timeIterPoints}
\end{subfigure}
\begin{subfigure}{0.5\textwidth}
\centering
\includegraphics[width=\textwidth]{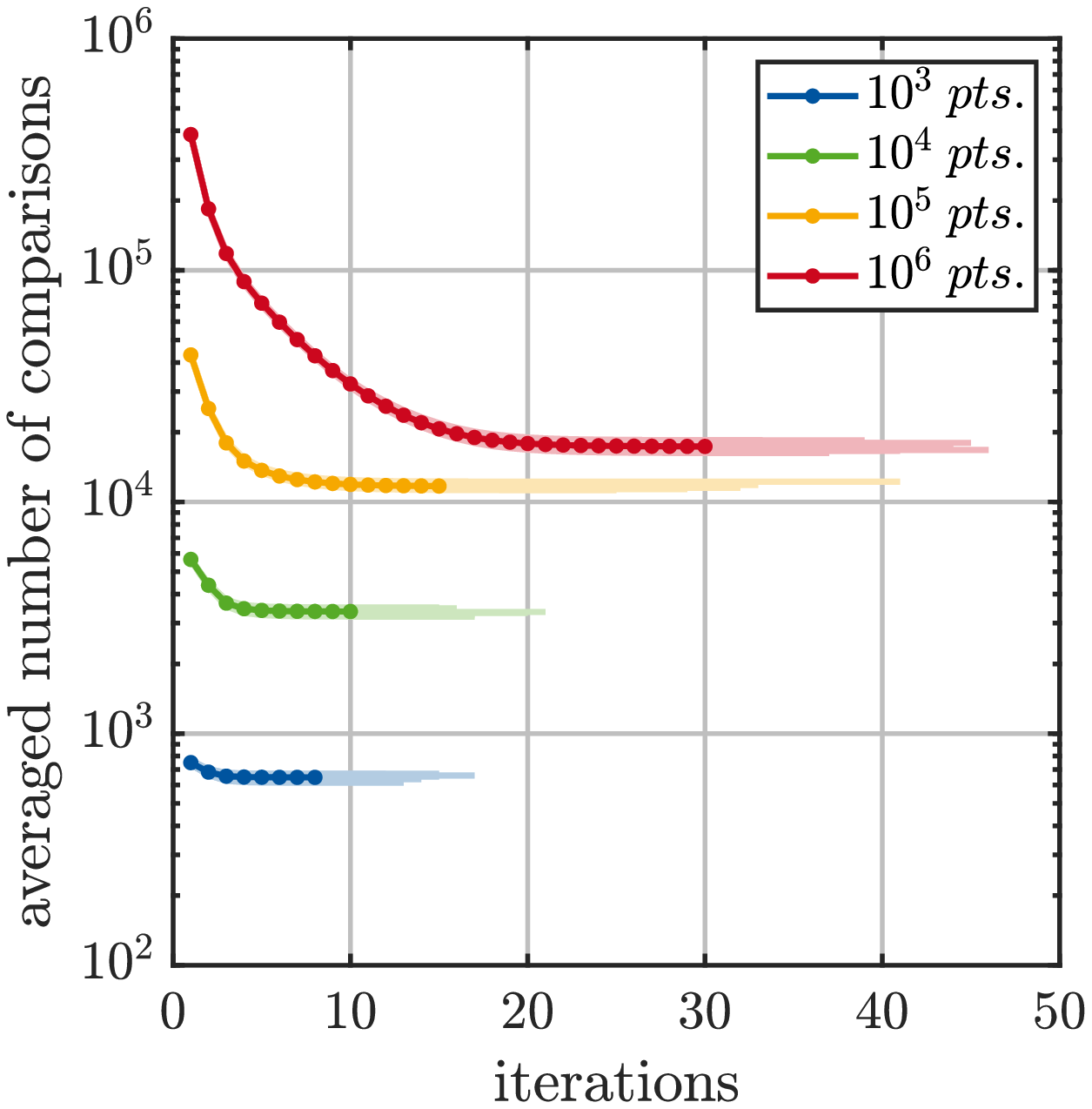}
\subcaption{}
\label{fig:compIterPoints}
\end{subfigure}
\caption{Elastic solid example: 12-d tree used as data structure for exact nearest-neighbor search. a) Averaged query times per iteration over iterations for different data set refinements. Averaged total query times stated in brackets. b) Distance comparisons per iteration over iterations for different data set refinements. Averaged results in bold, all results in light colors.}
\end{figure}

Initially, we compare the DD-solver's computation times for increasingly fine data sets of $10^3$, $10^4$, $10^5$, and $10^6$ points with 20 random samples each.
The studies were performed using a AMD Ryzen$^{\text{TM}}$ 9 3900X 12-core processor.
The c++ implementation makes use of the eigen3-library and the build-in Sparse-LU solver provided with this library.
The nearest-neighbor problem is solved using a 12-d tree based on an own implementation according to \cite{Bentley:1975, Friedman:1977}.
Times are measured for the assembly and LU decomposition of $\fK$, the right-hand side assembly of $\fE$ and $\fS$, solving the two equation systems (\ref{eq:KuE}) and (\ref{eq:KeS}), and finding the nearest-neighbors in the data sets.
The results of this study are depicted in Fig.~\ref{fig:timePoints}.
On the one hand, it can be observed that the total computational time increases heavily with the number of data points.
This is majorly caused by the fact that the time for the nearest-neighbor search increases drastically with the data set size.
On the other hand, with an increasing fine data set, the remaining squared distance decreases, as shown in Fig. \ref{fig:distIterPoints}.
The variance in the results reduces as well, which can be observed from the curves the light colors representing all performed computations.\\
Additional effects can be observed considering the computational times for assembling the right-hand sides and solving the two equation systems.
Here, the computational time for both parts slightly increases since the number of iterations needed to converge increases as well (see Fig.~\ref{fig:distIterPoints}).
The time for the initial assembly and LU decomposition of the stiffness matrix remains constant, as expected.
As an additional benchmark, we stated the time for computing the reference solution (6.2 s).
Here, the Newton-Raphson solver uses the ConjugateGradient solver from the eigen3 library.\\
From this observation we can conclude that the DD-solver's performance majorly depends on the time for solving the nearest-neighbor problem.
One can further remark that it might be reasonable to use more advanced and more problem-specific linear solvers than those used in this study.
Nevertheless, the study shows clearly that the crucial step remains the nearest-neighbor search.
Thus, the main focus is set on the investigation of the latter.\\
The critical reader might wonder why the times for finding the nearest-neighbors increase in the order of $O(N)$ and not in the order $O(log(N))$ as expected for $k$-d trees.
The reason for this effect can be detected if one studies the single search times per iteration.
These search times are depicted in Fig.~\ref{fig:timeIterPoints}.
For larger data sets, the main time effort is used in the first iterations.
For example, more than half ($51.6\%$) of the total search time is needed in the first three iterations of the $10^6$ points computation.
We can further study those observations regarding the number of distance comparisons that were made during the search.
The comparisons over iterations are shown in Fig.~\ref{fig:compIterPoints}.

\begin{figure}[htbp]
\centering
\includegraphics[width=\textwidth]{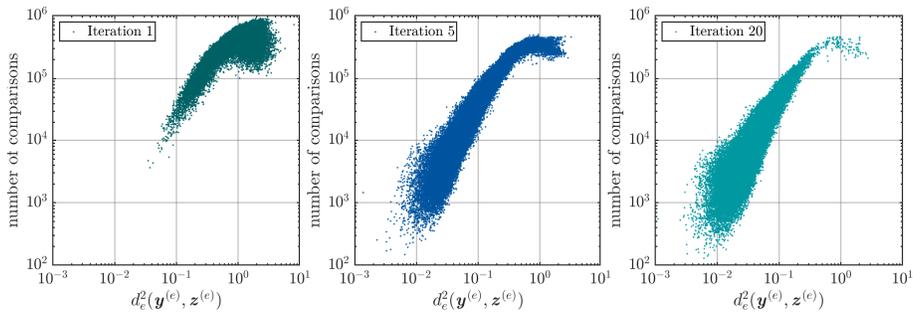}
\caption{Elastic solid example: number of comparisons over final query distance for all integration points in iteration 1 (left), 5 (mid) and 20 (right). Results of one computation with $10^6$ points.}
\label{fig:compDist}
\end{figure}

These results clearly correlate with the search times depicted in Fig.~\ref{fig:timeIterPoints}.
Another observation is that for the computations with $10^6$ points, on average approx.~40\% of all points have to be compared in the first iteration.
For smaller data sets the percentage even increases.
To further investigate this effect, we studied the correlation between the number of comparisons and the distances between the query point and its nearest-neighbor.
In Fig.~\ref{fig:compDist}, the number of needed comparisons is plotted over the final query distance of a million points sample in iteration 1, 5, and 20 for all integration points .
Here, the final query distances equal the final local distances $d_e^2(\fy\e,\fz\e)$.
Especially for iteration 5 and 20, a clear correlation between the distances and the number of comparisons can be observed.
For some query points in iteration 1, almost every point has to be checked.
A further observation is that many points move from the top right to the bottom left with increasing iterations.
That means the query distances decrease over the iterations, and as a consequence, the number of comparisons decreases as well.\\
Then finally, we have to clarify the reason for this correlation.
Therefore, we investigate the search in $k$-d trees in the following.
%

\subsection{Search in k-d trees}
\begin{figure}[htbp]
\begin{subfigure}{0.5\textwidth}
\centering
\psfrag{d}{$d_c$}
\boxed{\includegraphics[width=0.9\textwidth]{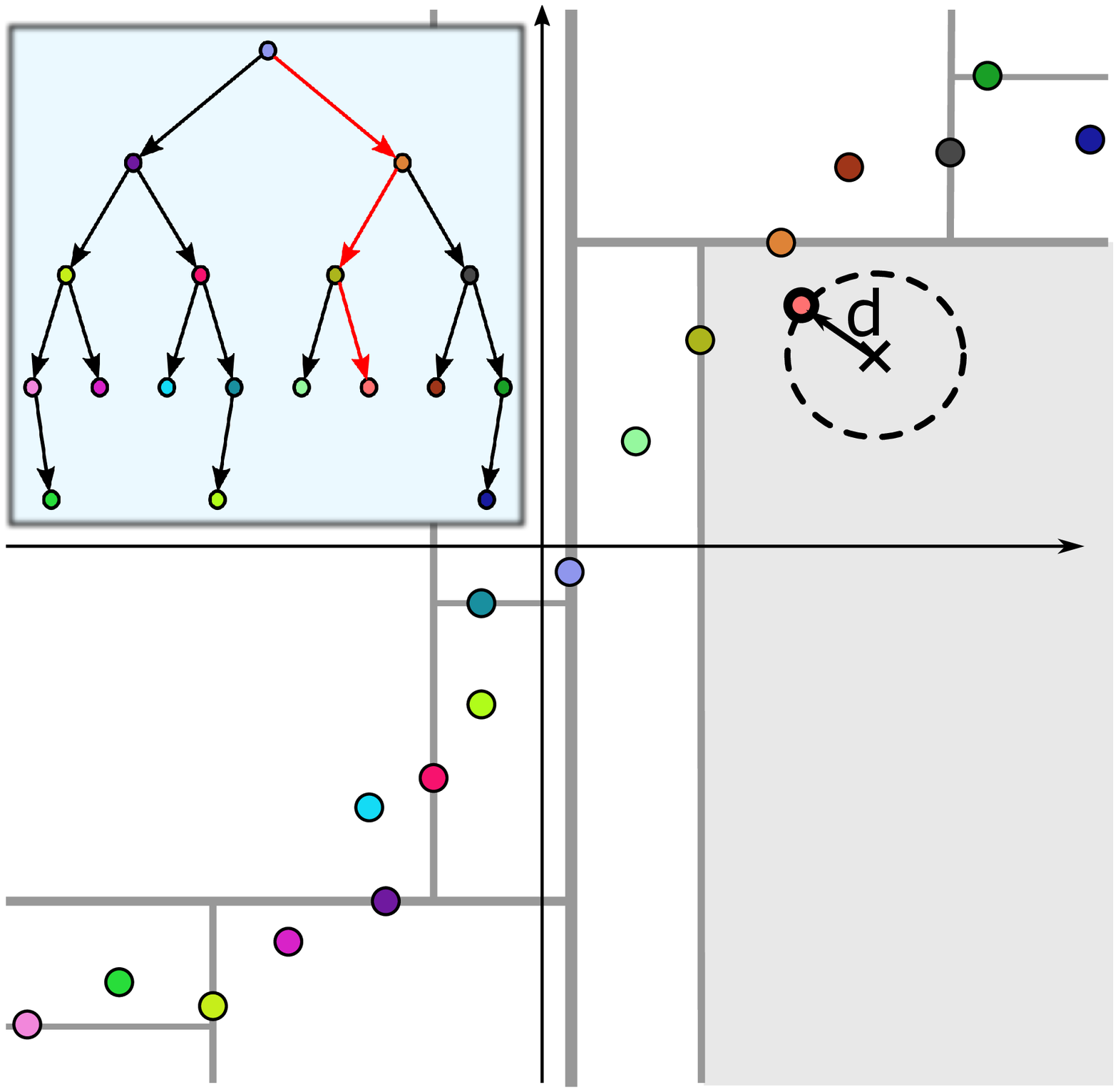}}
\subcaption{}
\label{fig:kdclose}
\end{subfigure}
\begin{subfigure}{0.5\textwidth}
\centering
\psfrag{d}{$d_c$}
\boxed{\includegraphics[width=0.9\textwidth]{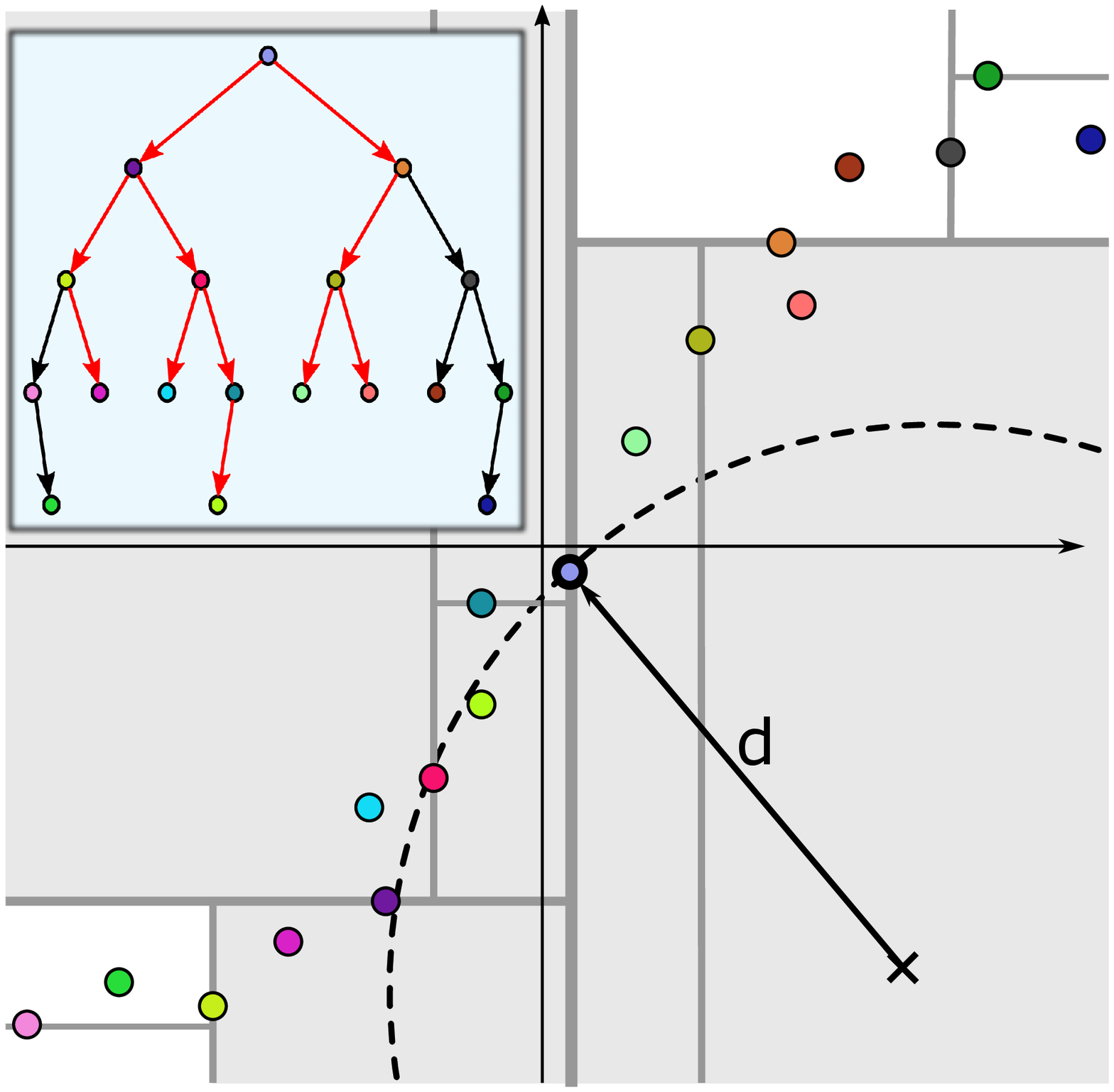}}
\subcaption{}
\label{fig:kdoff}
\end{subfigure}
\caption{Space partitioning by a $2$-d tree for a small data set with corresponding search tree. Sectors to be checked during the search for the closest point to the query point (black cross) are highlighted and corresponding search paths in the search trees are marked in red. Initial query distances $d_c$ indicated by dashed lines. a) Situation where the distance between the query point and its closest point is relatively small. b) Situation where the distance between the query point and its nearest-neighbor is rather large.}
\label{fig:kd}
\end{figure}

The construction of the $k$-d tree is based on the branch and bound method \cite{Land:2010}.
That means the data set is steadily divided into two branches according to the median coordinate of a certain branch dimension.
This median coordinate defines the bound of the branch.
Using this bound, additional checks in neighbored branches can be reduced.
Fig.~\ref{fig:kd} shows a tree of a small data set with 18 points created according to the above procedure.
The corresponding space partitioning is depicted as well. \\

The search in a $k$-d tree now works as follows.
Starting from the root node on top of the tree, the branch with its corresponding node is recursively chosen where the query point is positioned in.
The procedure ends when a chosen node has no children.
Thereupon, the backtracking starts, which means that all non-chosen branches are checked according to the defined bound.
As it is standard for $k$-d trees, non-chosen branches with according subbranches will be evaluated, if the following inequality is fulfilled:
\begin{equation}\label{eq:evalCriterion}
\underbrace{|x_i - q_i|}_{d_b} > d_c,
\end{equation}
where $x_i$ and $q_i$ are the coordinates in branch dimension $i$ of the branch node and the query node, respectively.
The absolute difference between both coordinates $d_b$ is then the distance to the corresponding branch bound, and $d_c$ represents the current best distance.
In terms of DD distances (see Eq. \ref{eq:distmetric}) the coordinates above are computed in an intermediate mapping where, e.~g., $\fq = (\C^{(e)1/2} \feps_i \;\;\; \C^{(e)-1/2}\fsig_i)^T$.\\

For material data in the phase space a special situation occurs, which is not typical for nearest-neighbor problems.
Since the material points typically lie in or close to a lower-dimensional manifold, two cases can be differentiated.
In the first situation, the query point is relatively close to the points in the data set.
The second situation deals with a query point which is far off the underlying manifold.
An example of situation 1 and situation 2 is explained in Fig.~\ref{fig:kdclose} and Fig.~\ref{fig:kdoff}, respectively.\\

\textit{Situation 1 (close query point):} First, we determine the sector where the query point is positioned in.
In the current example, this means only four comparisons are needed, which is advantageous compared to a linear search with 18 comparisons.
During backtracking, we find the situation that no additional branches have to be evaluated.
As illustrated in Fig.~\ref{fig:kdclose}, figuratively speaking, this is because there is no intersection between a voxel boundary and the circle around the query point with the best distance radius.
This means in the current example that the use of a 2-d tree would be advantageous compared to a linear search with 18 comparisons.\\

\textit{Situation 2 (far off query point):} Again, we start following the same voxel's query path as in the above situation.
Here, the light purple root node is the current best point so far (see Fig.~\ref{fig:kdoff}).
Now, there is the situation that numerous intersections between the dashed black line and the voxel boundaries exist.
Thus, we have to investigate additional branches, and more comparisons have to be made.
Finally, we compare 12 out of 18 nodes.
All checked sectors are highlighted, and the corresponding children in the search tree are marked in red.
For this higher ratio of comparisons, the 2-d tree might even be dominated by a linear search due to its additional overhead. \\

This simple example gives an explanation for the correlation observed in Fig.~\ref{fig:compDist}.
The effect intensifies for higher dimensional spaces because here, the distances between the points increase as well  (see e.~g. \cite{Uhlmann:1991, Sproull:1991}).
%
%
\section{Nearest neighbor approximations}
The previous chapter showed that solving the nearest-neighbor problem dominates the whole solution procedure for larger data sets.
Therefore, we compared different approximate nearest-neighbor algorithms to speed up the search.
In all presented approximation algorithms, the trade-off between computational efficiency and accuracy will be investigated.\\
As already mentioned in the introduction, the first central idea of this work is to control the accuracy so that it increases during the data-driven solver's iterations.
If we remember that the nearest-neighbors found in the first iterations are only intermediate results, we can assume that it will be sufficient to demand only a low accuracy at the beginning.
In the very first iterations, a very rough guess might even be good enough.
That should lead to high computational time savings since those will be the most time-intensive ones, at least if we use a $k$-d tree. \\
A simple implementation could e.~g. control an arbitrary search parameter $f$ which influences the accuracy
\begin{equation}
	f = \hat f(d(\fz,\fy),d_e(\fz\e,\fy\e),i),
\end{equation}
where the parameter could be a function of the latest global or local distance and the iteration number $i$.\\
%

\subsection{Approximate nearest-neighbors using k-d trees}\label{sec:approxkd}
\begin{figure}[!htbp]
\begin{subfigure}{0.5\textwidth}
\centering
\includegraphics[width=0.8\textwidth]{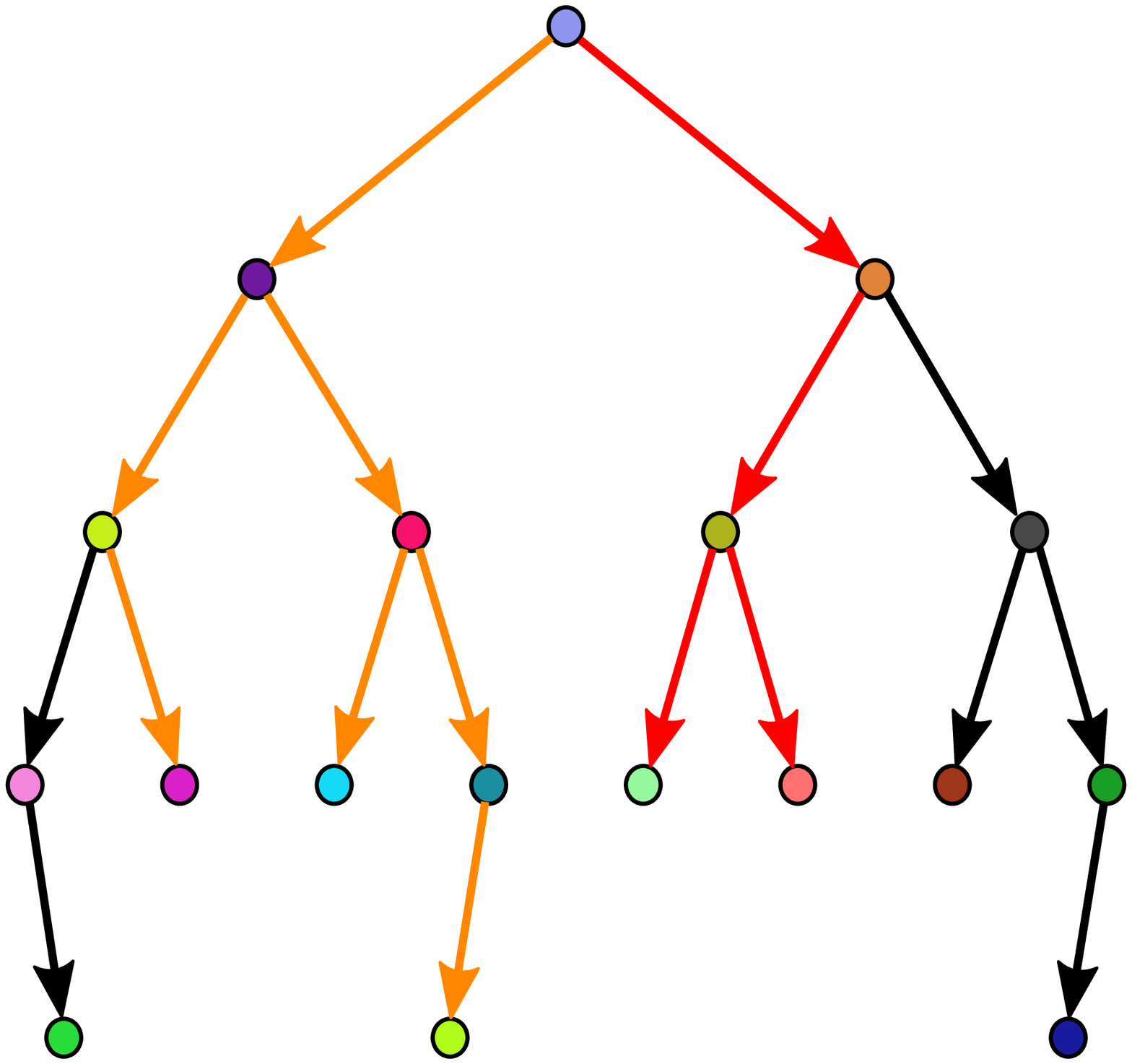}
\subcaption{}
\label{fig:evalReduction}
\end{subfigure}
\begin{subfigure}{0.5\textwidth}
\centering
\psfrag{d}{$d_c$}
\psfrag{fd}{$f_d\!\cdot\! d_c$}
\psfrag{x}{$d_b$}
\includegraphics[width=0.9\textwidth]{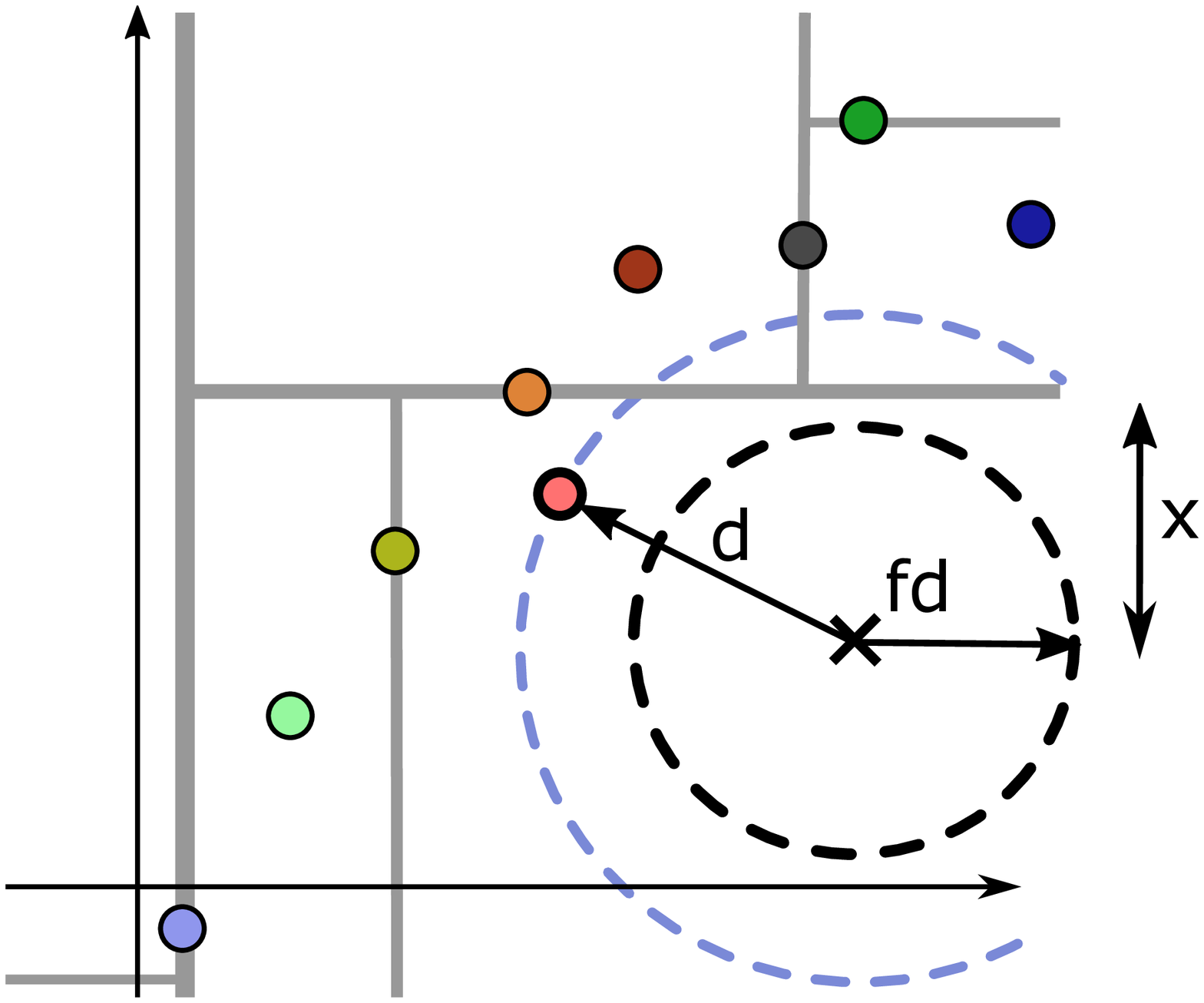}
\subcaption{}
\label{fig:distReduction}
\end{subfigure}
\caption{a) Illustration of reduced search paths in a tree structure. Red paths are evaluated, whereas the orange paths are neglected. b) Reduction of the number of checks by reducing the current best distance $d_c$ by a factor $f_d$ for backtracking.}
\end{figure}

All tree-based approximate nearest-neighbor algorithms have in common that they reduce additional branch evaluations during backtracking.
 %
For example, Fig.~\ref{fig:evalReduction} shows a search tree for a specific query request.
In this case, we perform a single run from top to bottom so that only the red paths are evaluated.
To perform an exact search, where we can ensure to find the exact nearest-neighbor, the orange paths would have to be evaluated in addition.
By neglecting the orange paths during backtracking, we accelerate the search because less distance comparisons have to be done.
In contrast, the accuracy reduces since we might find a point which is just close, but not the closest.
There are several ways to achieve this.
One possibility is to limit the number of additional evaluations during backtracking.
In \cite{Silpa:2008} this procedure is investigated with the additional use of a priority queue on multiple randomized $k$-d trees.
Several differently partitioned $k$-d trees are assessed, and those non-chosen branches closest to the query point are evaluated first.
In \cite{Muja:2009a}, the authors compare this procedure to different methods and provide an implementation in the FLANN-library \cite{Muja:2009b}.\\
Another procedure, which we followed here, is proposed in \cite{Arya:1998}.
The idea is to introduce a control parameter $f_d \in [0,1] $, which is multiplied with the current best distance $d_c$ so that inequality \ref{eq:evalCriterion} is modified to
\begin{equation}
d_b > f_d \cdot d_c.
\end{equation}
If we chose $f_d$ to be one, the exact $k$-d tree will be recovered.
Counter-wise, if $f_d=0$, no additional branch will be evaluated.
Thus, additional evaluations can be reduced because the number of bound intersections is decreased (see Fig.~\ref{fig:distReduction}).\\

\begin{figure}[!htbp]
\begin{subfigure}{0.5\textwidth}
\centering
\includegraphics[width=\textwidth]{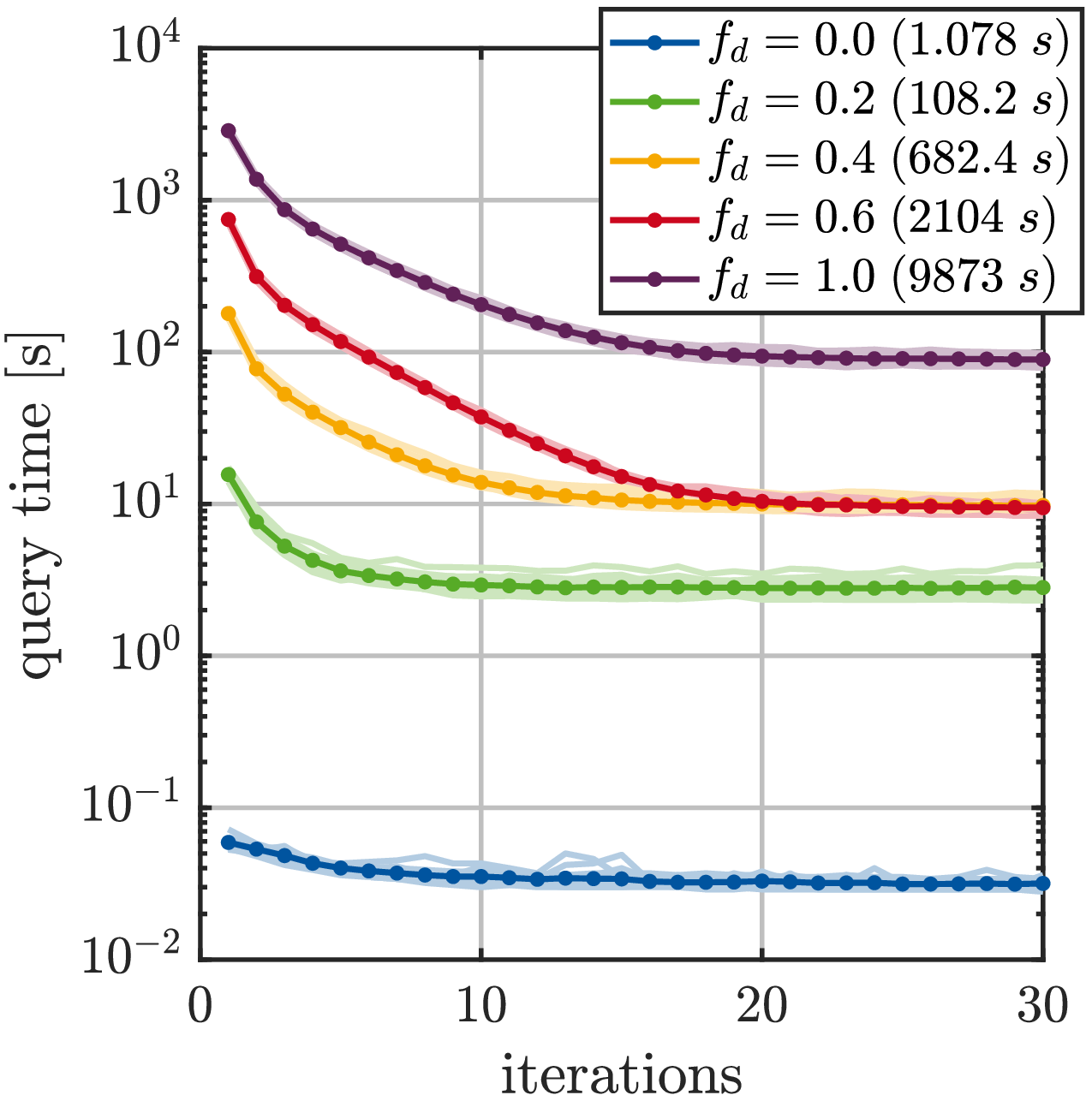}
\subcaption{}
\label{fig:timeIterkdfd}
\end{subfigure}
\begin{subfigure}{0.5\textwidth}
\centering
\includegraphics[width=\textwidth]{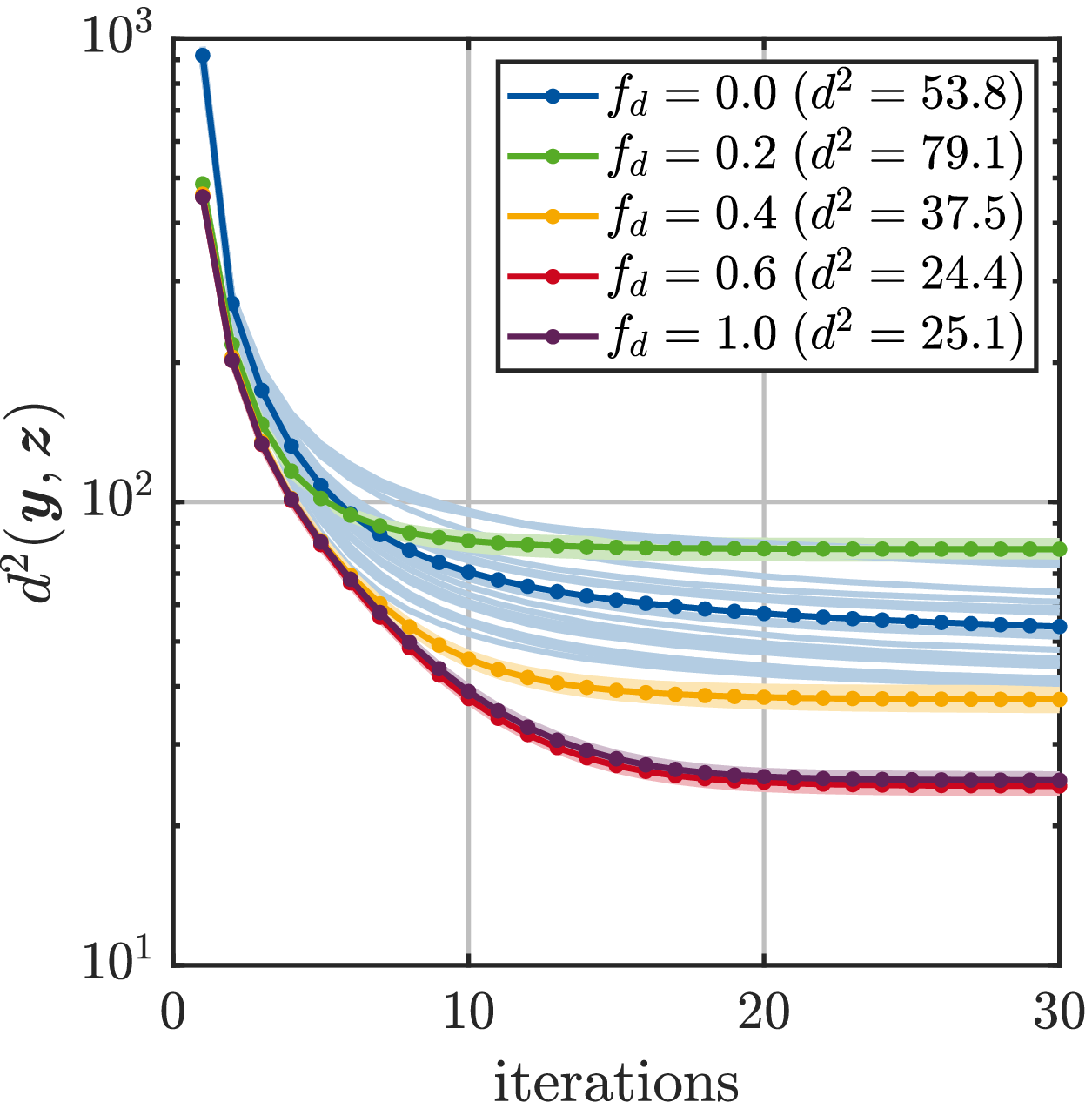}
\subcaption{}
\label{fig:distIterkdfd}
\end{subfigure}
\caption{Elastic solid example: Approximate $12$-d tree used for nearest-neighbor search on 20 one million points data sets. Results are studied for different parameters $f_d$. Averaged results are depicted in bold, all results in light colors. a) Query times per iteration over solver iterations. Total search times within the first 30 iterations stated in brackets. b) Global squared distance over solver iterations. Remaining squared distance after 30 iterations stated in brackets.}
\end{figure}

Parameter studies were performed for the described implementation of the $k$-d tree with parameters $f_d=0,0.2,0.4,0.6,1.0$.
The same 20 samples of the 1 million points data sets, which we already used in the study of section \ref{sec:exactkd}, were investigated.
These results are depicted in Fig.~\ref{fig:timeIterkdfd} and Fig.~\ref{fig:distIterkdfd} showing the query times and distances over the iterations.
For the sake of comparability, we consider the first 30 iterations only.\\
On the one hand, we can reduce the search times per iteration by almost 4 magnitudes.
As mentioned before, for $f_d=1.0$ the same results as for the exact $k$-d tree in Fig.~\ref{fig:kD_points} are retrieved.
For the control parameter $f_d = 0.0$, we perceive query times of less than 0.04 seconds per iteration on average.
Regarding the query times of the 20 samples, only very small fluctuations occur.\\
On the other hand, in Fig.~\ref{fig:distIterkdfd}, we observe that the remaining distances after 30 iterations increase with decreasing values of $f_d$.
But two exceptions from this can be seen.
First, the remaining distances with $f_d=0.0$ fall below those with $f_d=0.2$.
Second, the results with $f_d=0.6$ show on average slightly lower distances than the computations with $f_d=1.0$.\\

To explain these observations is difficult, but the authors have the following assumption.
As mentioned above, the data-driven iterative solver might stop in local minima.
The worse the nearest-neighbor algorithm's accuracy, the higher the probability of running into a local minimum.
However, a very ``bad guess'' for one integration point in an iteration can lead to a better solution if a larger step beyond a local minimum is done in the following iteration.
This assumption would also explain the high variability observed for the computations with $f_d=0.0$ because with lower $f_d$-values more local minima exist and larger local steps are more likely.
The described effect seems to be intensified due to the $k$-d space partitioning, as we can conclude from the following section.
%

%
\subsection{Approximate nearest-neighbors using k-means trees}\label{sec:kmeans}
As a second tree-based search algorithm, we investigate the $k$-means tree.
This tree was proposed in the early work of Fukunga \& Narendra \cite{Fukunaga:1975}.
It is based on the idea of recursively branching the data into $k$ clusters until all clusters are smaller than a certain bucket size.
This clustering can be performed in principle by different clustering techniques.
In \cite{Fukunaga:1975}, the $k$-means clustering algorithm was suggested, which we also used in our studies.
Then, the node corresponding to a cluster bears the coordinates of the midpoint of all points in this cluster.
The node is also equipped with the largest distance $d_r$ from the cluster mean to all points included in the cluster.\\

The search in the $k$-means tree starts from the root node.
Then, recursively all distances to its children are computed and sorted.
The algorithm then chooses the child or cluster closest to the query point as the next node to evaluate until a leaf node is reached.
During backtracking the non-chosen clusters are evaluated if
\begin{equation}
	d(\fx,\fq) - f_d \cdot d_r < d_c,
\end{equation}
where $d(\fx,\fq)$ defines the distance between the non-chosen cluster midpoint $\fx$ and the query point $\fq$.
Here, the triangular inequality is exploited to reduce the number of necessary evaluations.
Besides, the accuracy control parameter $f_d$ is used in a similar manner as introduced before in section \ref{sec:approxkd}.
However, here $f_d$ reduces the radius $d_r$ to ensure that the exact search is performed for $f_d=1.0$ and no backtracking is done for $f_d=0.0$.\\

\begin{figure}[!htbp]
\begin{subfigure}{0.5\textwidth}
\centering
\includegraphics[width=0.9\textwidth]{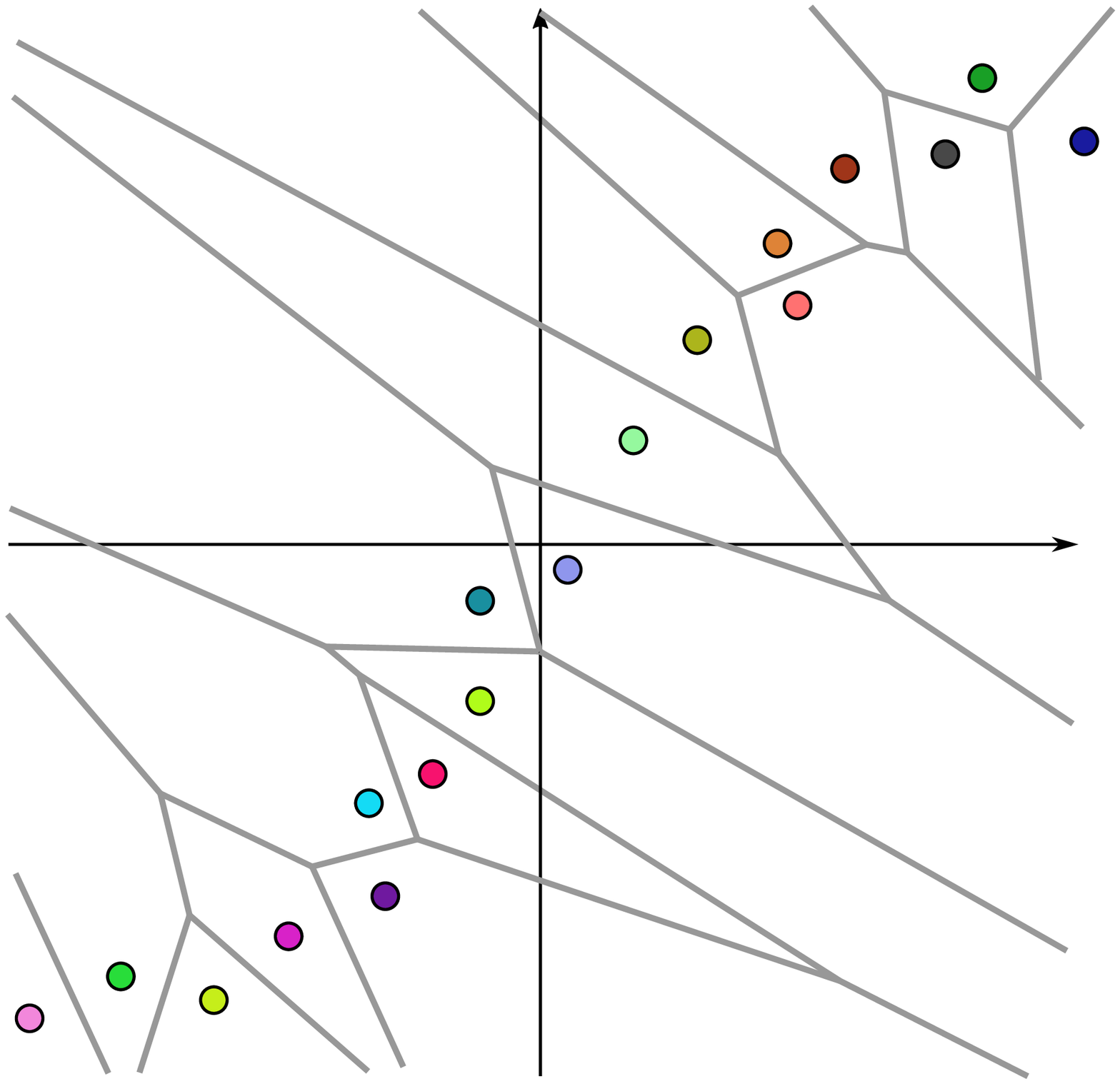}
\subcaption{}
\label{fig:voronoi}
\end{subfigure}
\begin{subfigure}{0.5\textwidth}
\centering
\psfrag{d}{$d_c$}
\includegraphics[width=0.9\textwidth]{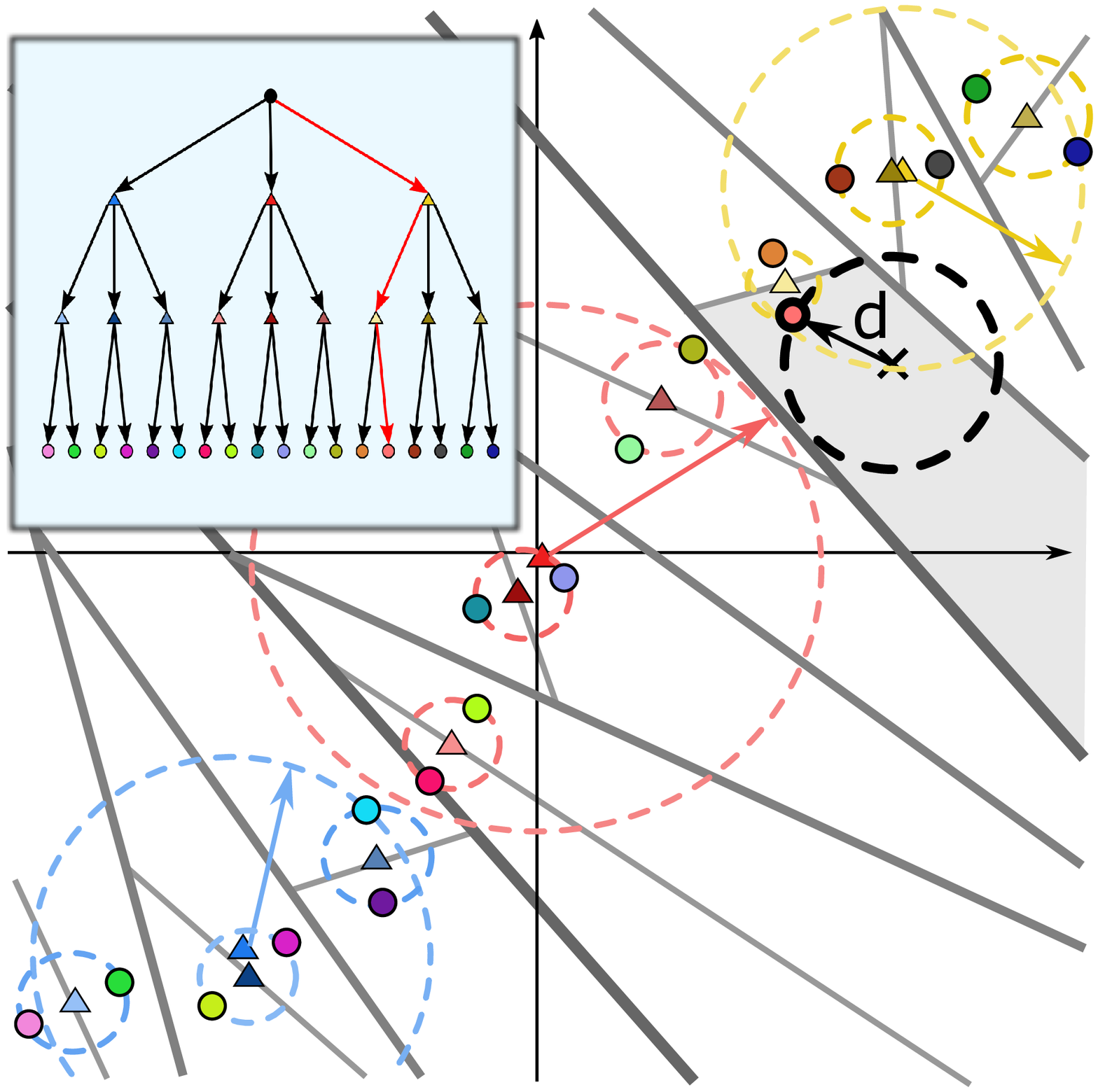}
\subcaption{}
\label{fig:kmeanstree}
\end{subfigure}
\caption{a.) Voronoi tesselation of a small data set. b) Space partitioning by a $3$-means tree for the same data set with corresponding search tree. Sectors to be checked during the closest-point search to the query point (black cross) are highlighted and corresponding search paths in the search tree are marked in red. Final query distance indicated by black dashed line. Center points of clusters are indicated by triangles. Relevant cluster radii $d_r$ for back checks are marked by colored dashed lines.}
\end{figure}

According to the described $k$-means tree, a space partitioning is depicted in Fig.~\ref{fig:kmeanstree} with $k=3$.
First of all, the space partitioning is now similar to the data set's Voronoi tessellation (see Fig.~\ref{fig:voronoi}).
This space partitioning appears to be more natural to the problem compared to the before discussed $2$-d partitioning.
It is here sufficient to run a single search path to find the exact result regarding the depicted query situation.
This is because there is no intersection between the circle around the query point with radius $d_c$ and the circle around any non-evaluated cluster with radii $d_r$.

\begin{figure}[htbp]
\begin{subfigure}{0.5\textwidth}
\centering
\includegraphics[width=\textwidth]{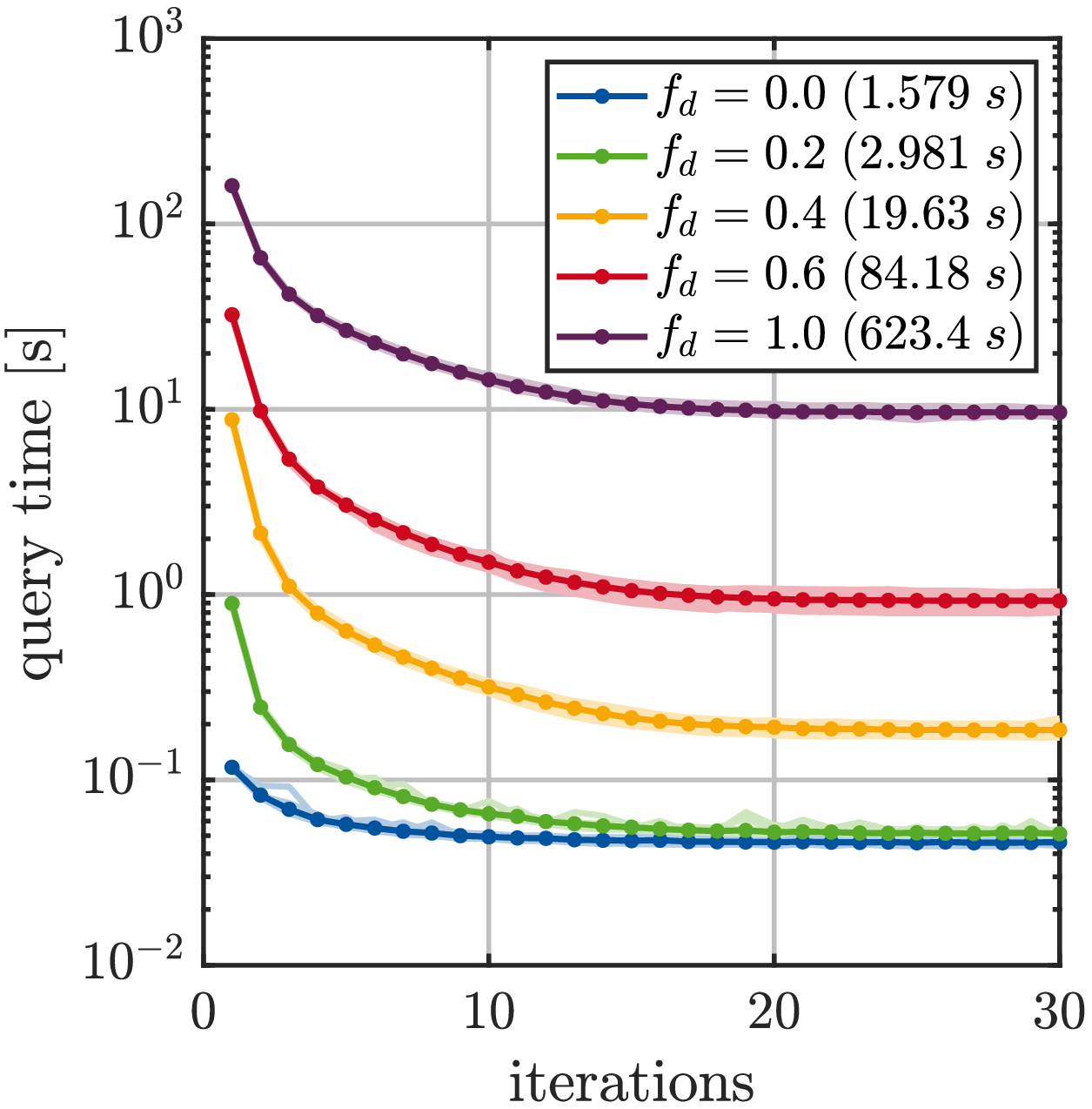}
\subcaption{}
\label{fig:timeIterkmeans}
\end{subfigure}
\begin{subfigure}{0.5\textwidth}
\centering
\includegraphics[width=\textwidth]{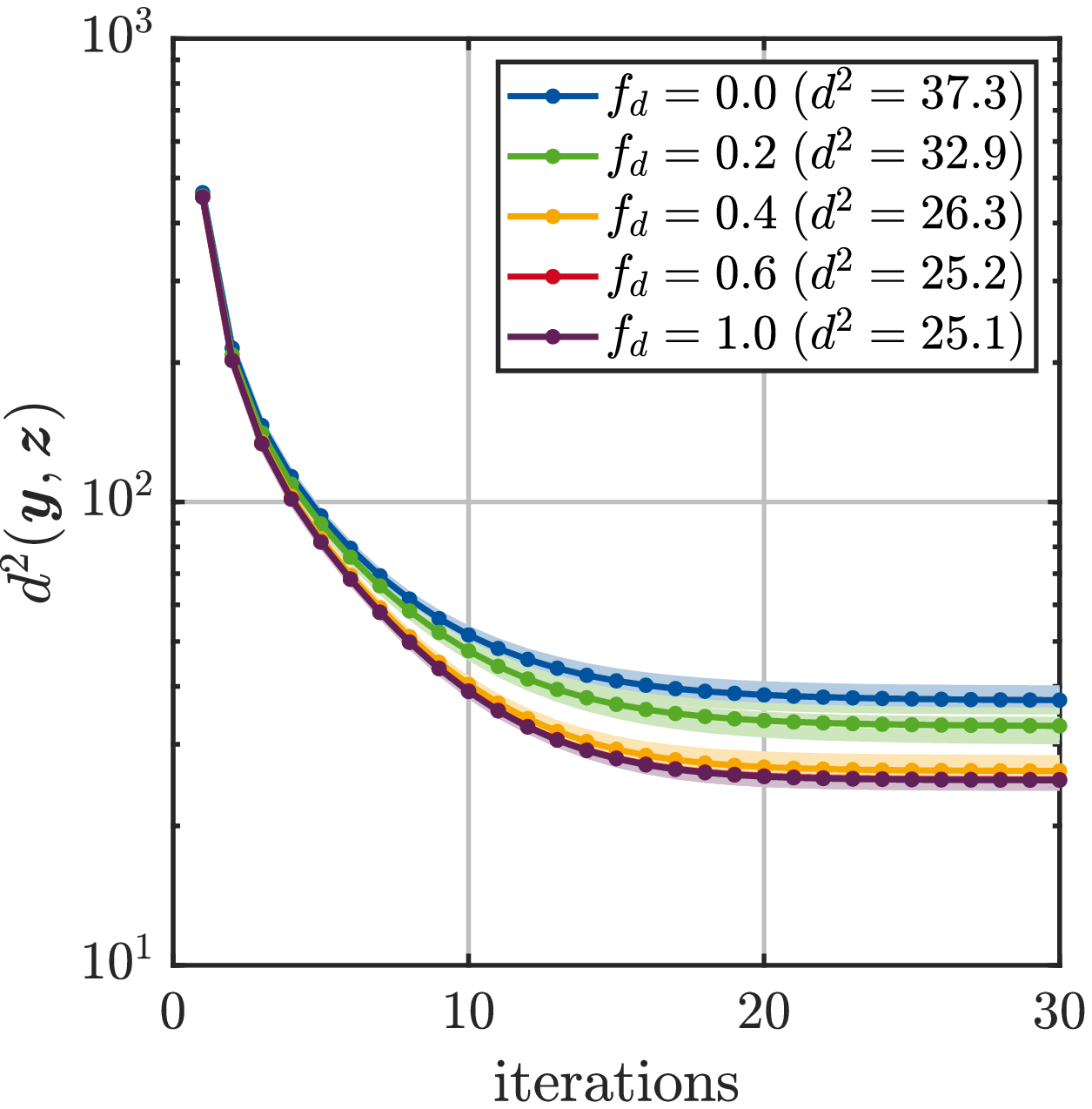}
\subcaption{}
\label{fig:compIterkmeans}
\end{subfigure}
\caption{Elastic solid example: Approximate $4$-means tree used for nearest-neighbor search on 20 one million points data sets. Results are studied for different parameters $f_d$. Averaged results are depicted in bold, all results in light colors. a) Query times per iteration over solver iterations. Total search times within the first 30 iterations stated in brackets. b) Global squared distance over solver iterations. Remaining squared distance after 30 iterations stated in brackets.}
\end{figure}

A parameter study according to the same procedure as in the section before was performed with values $f_d = 0.0, 0.2, 0.4, 0.6, 1.0$.
Therefore, again the same 20 samples with one million points were used.
Here, a $4$-means tree was investigated with a bucket size of $k^2=16$.
The choice of $k=4$ was made here since the query times of studies with $k=2$ showed similar results, but the needed memory is lower.
For higher values of $k$, the query times significantly increased.
The results of these studies are depicted in Fig.~\ref{fig:timeIterkmeans} and Fig.~\ref{fig:compIterkmeans}.\\

Interestingly, the query times for the exact cases with $f_d=1.0$ are about one magnitude smaller compared to those computed with the $k$-d tree.
In contrast to this, the query times for $f_d = 0.0$ remain approximately at the same speed.
However, for the computations with $f_d=0.2$ and $f_d=0.4$ the query times are relatively low.\\
Concentrating on the accuracy, already for $f_d=0.4$ an excellent behavior is seen.
Even the results with $f_d=0.0$ are quite accurate and show only a small variation.

A further remark regarding the comparability to open-source algorithms should be added.
We performed the same tests with the implementations of the FLANN-library and found that the results were comparable.
However, our implementations slightly dominated those results regarding query time and offered more insight.
Besides, in the authors' opinion controlling accuracy is more intuitive in the way discussed above.
%

\subsection{Approximate nearest-neighbors using k-nearest-neighbor graphs}
\begin{figure}[!htbp]
\centering
\psfrag{s1}{$\boldsymbol s_1$}
\psfrag{s2}{$\boldsymbol p_1 = \boldsymbol s_2$}
\psfrag{q1}{$\fq_1$}
\psfrag{q2}{$\fq_2$}
\psfrag{x}{$\fp_2$}
\includegraphics[width=0.6\textwidth]{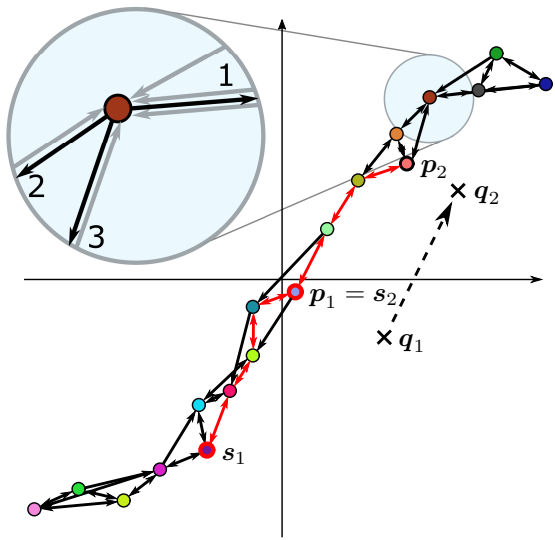}
\caption{Graph structure of a small data set using a 3-nearest-neighbor graph (see enlargement). Query path from a starting point $\fs_1$ to the closest point $\fp_1$ of the query point $\fq_1$ is highlighted in red. Closest point $\fp_1$ is new starting point $\fs_2$ for query in second iteration.}
\label{fig:graphsearch}
\end{figure}

Recently the category of graph searches has won much attention.
For higher-dimensional data sets, this type of search algorithm seems to be superior.
Therefore, the investigation of the latter is of the highest interest.\\
Here, we investigate the search in a $k$-NN graph initially introduced in \cite{Hajebi:2011}.
In contrast to the tree search algorithms discussed before, graph search algorithms are not based on branch and bound algorithms.
Instead, every point in the data set is connected with a couple of other points from the same set.
Considering the $k$-NN graph, a network is built by connecting every node to its $k$ nearest-neighbors.
In this manner, a network is created as depicted in Fig.~\ref{fig:graphsearch}, where every point is connected with its $k=3$ nearest-neighbors.
The probably simplest strategy is then to choose a random starting point and use a greedy algorithm to find the nearest point.
That means recursively, all distances between the neighbors of the current point and the query points are computed, and this point closest to the query point is selected as the next current.
If the current point itself is the closest, the search will stop and the current point will be identified as the nearest-neighbor.
For sure, this procedure will stop in a local minimum if the number of nearest-neighbors $k$ in the network is set too small.
This number $k$ also has a large impact on the graph's properties like the sparsity, connectivity, or search time.
Therefore, the accuracy can be controlled, e.~g., by increasing or decreasing the number of nearest-neighbors or searching multiple times with different starting points.\\

\begin{figure}[!htbp]
\begin{subfigure}{0.5\textwidth}
\centering
\includegraphics[width=\textwidth]{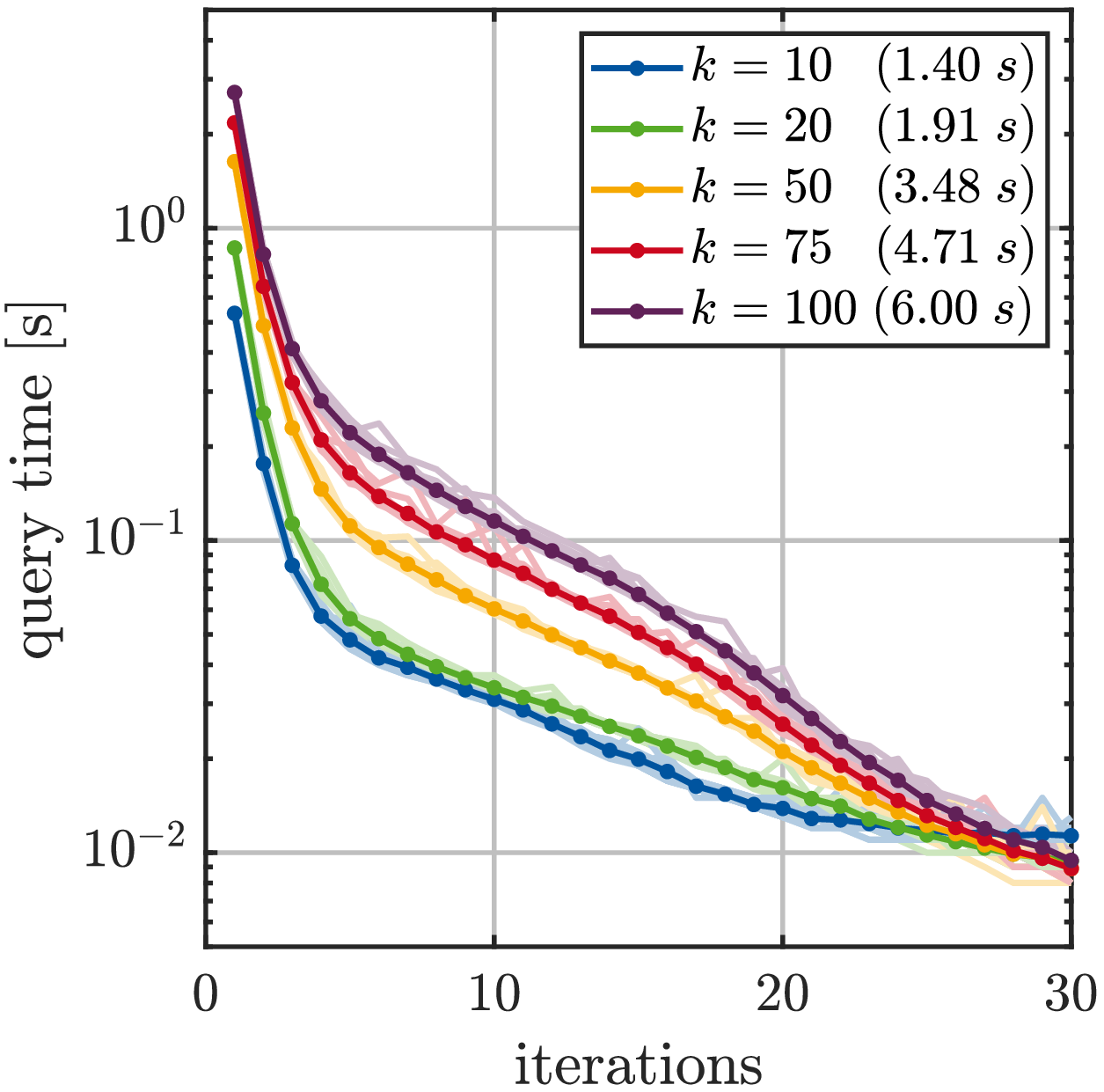}
\subcaption{}
\label{fig:timeIterkNNG}
\end{subfigure}
\begin{subfigure}{0.5\textwidth}
\centering
\includegraphics[width=\textwidth]{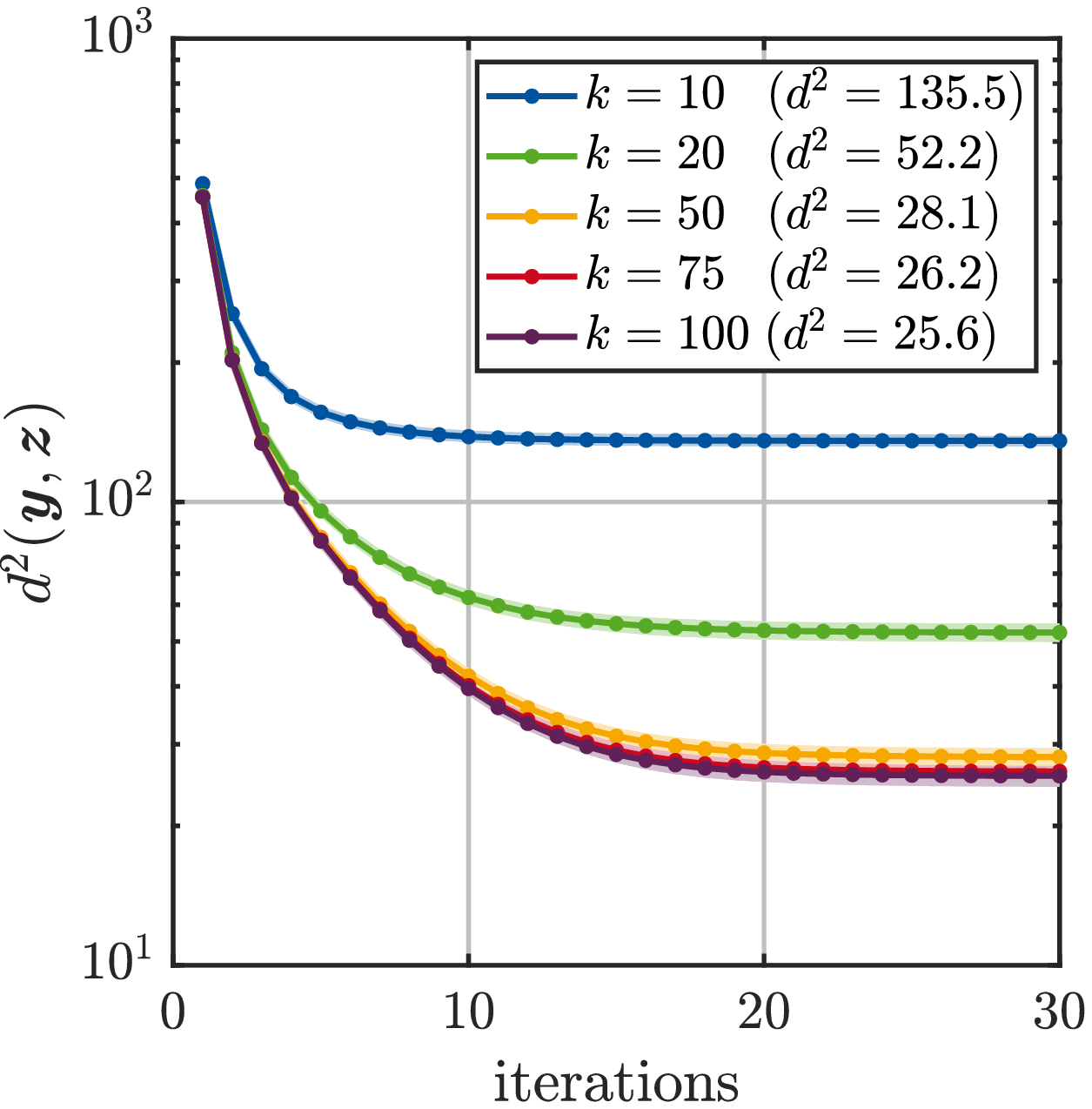}
\subcaption{}
\label{fig:distIterkNNG}
\end{subfigure}
\caption{Elastic solid example: Approximate $k$-NN graph used for nearest-neighbor search on 20 one million points data sets. Results are studied for different parameters $k$. Averaged results are depicted in bold, all results in light colors. a) Query times per iteration over solver iterations. Total search times within the first 30 iterations stated in brackets. b) Global squared distance over solver iterations. Remaining squared distance after 30 iterations stated in brackets.}
\end{figure}

This leads to the second central point regarding the combination of search algorithms and the data-driven solver.
Since we observe convergence towards a solution with increasing iterations, the single query points $\fy\e$ will move less and less.
This effect can be exemplarily seen in Fig.~\ref{fig:projections}.
Therefore, the knowledge about the nearest-neighbor of the previous iteration can be used.
A simple approach is to select the previous solution as the starting point for the next iteration, as illustrated in Fig.~\ref{fig:graphsearch}.
With ongoing iterations, the search path length should then decrease and thus also the query time.\\
The presented problem of a moving query point is also addressed, e.~g., in \cite{Song:2001,Guting:2010,Yu:2014}.
Here, the problem of nearest-neighbor queries is applied to moving mobile devices in the 3D space.
In \cite{Song:2001}, the authors prove that no additional search has to be performed if a query point moves less than a small distance $\delta$.
In our context, the relation can be estimated to
\begin{equation}
 	d(\fq_i,\fq_{i+1}) < \frac{d(\fq_i,\fp_{i,1})-d(\fq_i,\fp_{i,2})}{2} = \delta,
\end{equation}
where $\fq_i$ and $\fq_{i+1}$ are the query points in iteration $i$ and $i+1$, respectively.
Further, $\fp_{i,1}$ and $\fp_{i,2}$ are the closest and second closest points, respectively, from the search in iteration $i$.
Then, $\delta$ defines the maximum distance a query point can move from one iteration to another so that no query has to be performed.\\
We studied the described graph search algorithm's behavior for its parameter $k$ on the same 20 samples of the one million points data sets as used before.
The neighborhood sizes were varied to $k= 10,20,50,75,100$.
The graph structures were built using the approximate $4$-means tree with $f_d=0.6$, discussed in the previous section.
The results of these studies are depicted in Fig.~\ref{fig:timeIterkNNG} and Fig.~\ref{fig:distIterkNNG}.\\
\begin{figure}[!htbp]
\begin{subfigure}{0.5\textwidth}
\centering
\includegraphics[width=\textwidth]{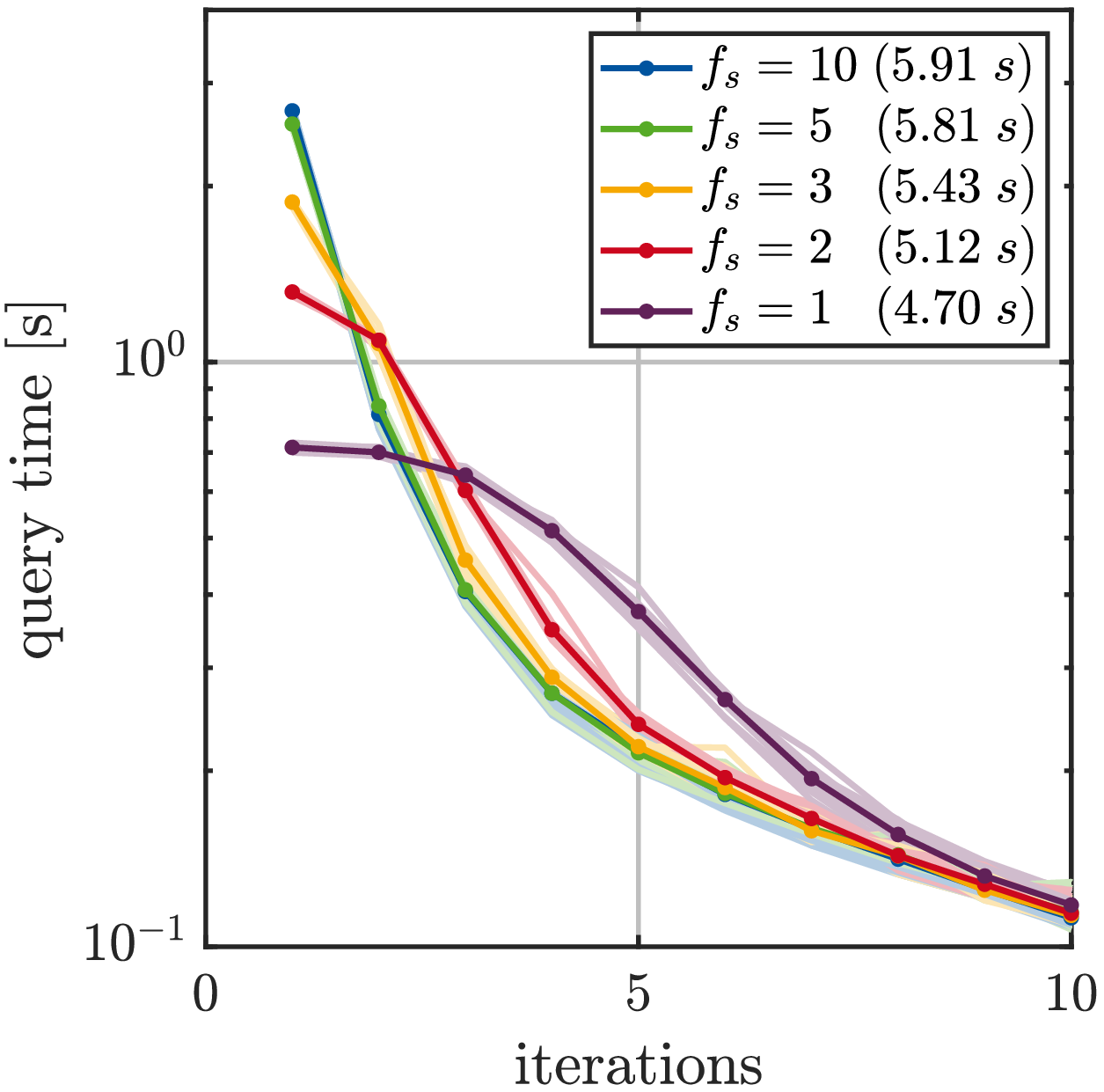}
\subcaption{}
\label{fig:timeIterkNNGsteps}
\end{subfigure}
\begin{subfigure}{0.5\textwidth}
\centering
\includegraphics[width=\textwidth]{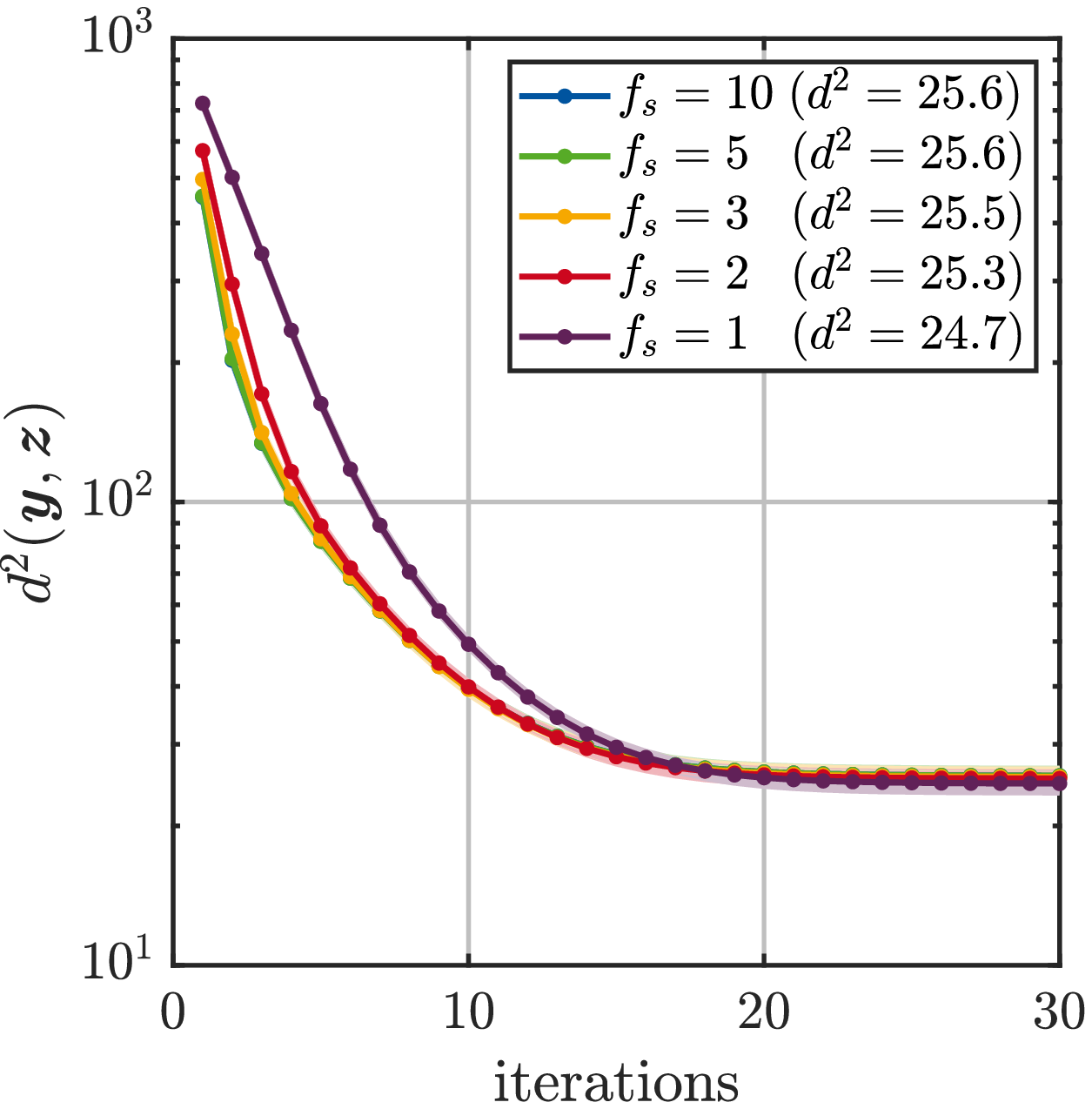}
\subcaption{}
\label{fig:distIterkNNGsteps}
\end{subfigure}
\caption{Elastic solid example: Approximate $k$-NN graph used for nearest-neighbor search on 20 one million points data sets. Results are studied for different parameters $f_s$. Averaged results are depicted in bold, all results in light colors. a) Query times per iteration over solver iterations. Total search times within the first 30 iterations stated in brackets. b) Global squared distance over solver iterations. Remaining squared distance after 30 iterations stated in brackets.}
\end{figure}
Regarding the query times, a decrease of almost two amplitudes can be observed from iteration one to thirty for all parameter configurations.
We further saw that the lower the neighborhood size $k$, the lower are also the needed search times.
However, this goes along with a significant loss of accuracy, especially for $k=10$ and $k=20$.
Increasing the neighborhood size from $k=75$ to $k=100$, almost no benefit regarding efficiency shows up.\\
The $k$-NN graph's results discussed above are already quite promising, but still we see relatively long query times in the first iterations.
Therefore, we performed an additional study where we limited the number of node changes to a bound $f_s$.
Hereby, we aimed to reduce the number of comparisons in the first iterations, where the search paths are longest.
The computations were performed with bounds $f_s = 10,5,3,2,1$.\\
In Fig.~\ref{fig:timeIterkNNGsteps}, the query times for the first ten iterations are depicted.
As intended, the query times in the first iteration can be reduced, whereas the query times in the following are slightly larger than before.
In total, the query times can be reduced by more than 20\% for $f_s=1$ compared to $f_s=10$.
Interestingly, the remaining distances after 30 iterations for $f_s=1$ slightly dominate the other computations (see. Fig.~\ref{fig:distIterkNNGsteps}).
Further, the smaller $f_s$, the more gradual decreases the distance.
%

\subsection{Comparisons of different algorithms on a 100 million points data set}
\begin{figure}[!htbp]
\begin{subfigure}{0.5\textwidth}
\centering
\includegraphics[width=\textwidth]{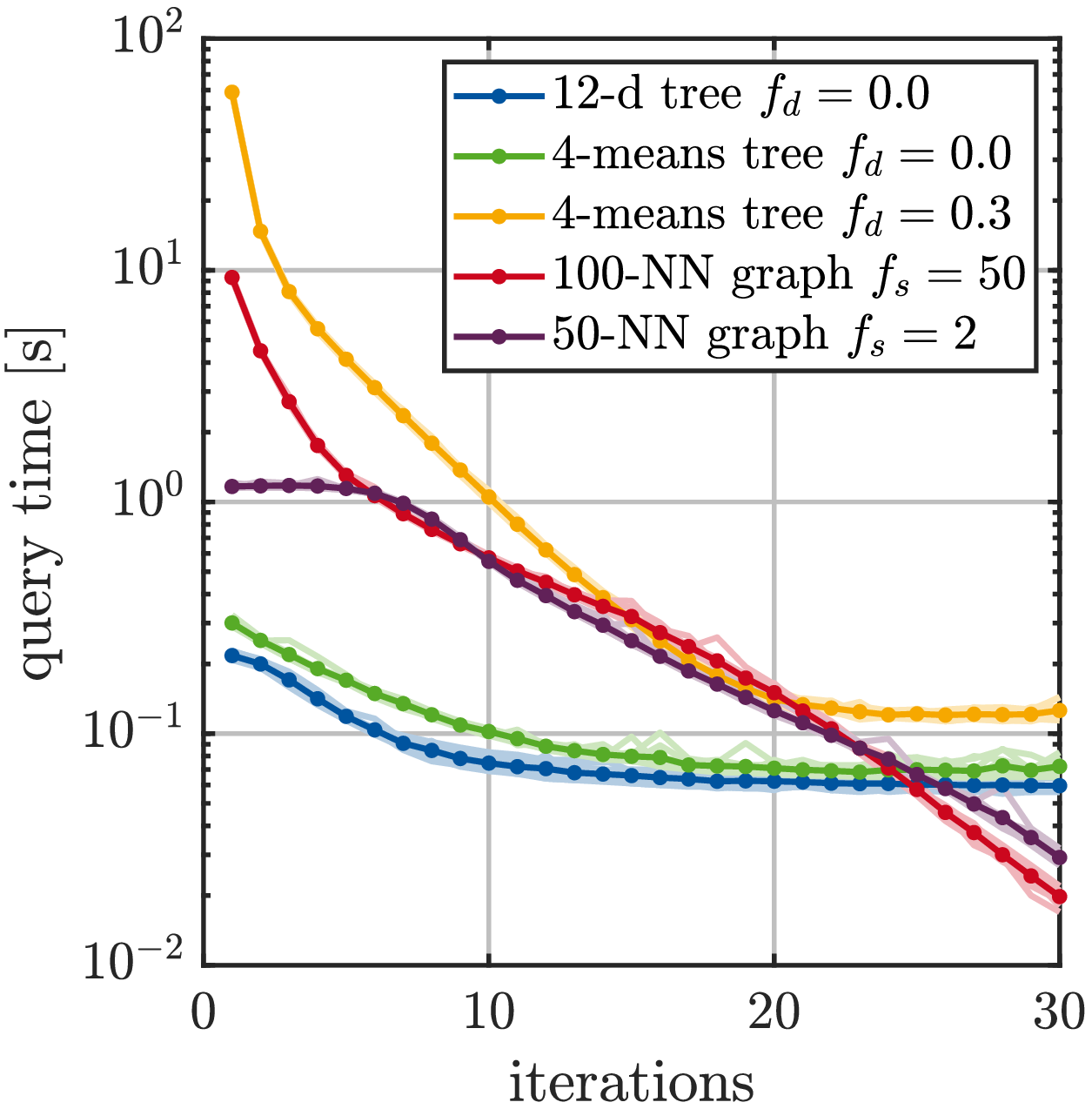}
\subcaption{}
\label{fig:timeIter100million}
\end{subfigure}
\begin{subfigure}{0.5\textwidth}
\centering
\includegraphics[width=\textwidth]{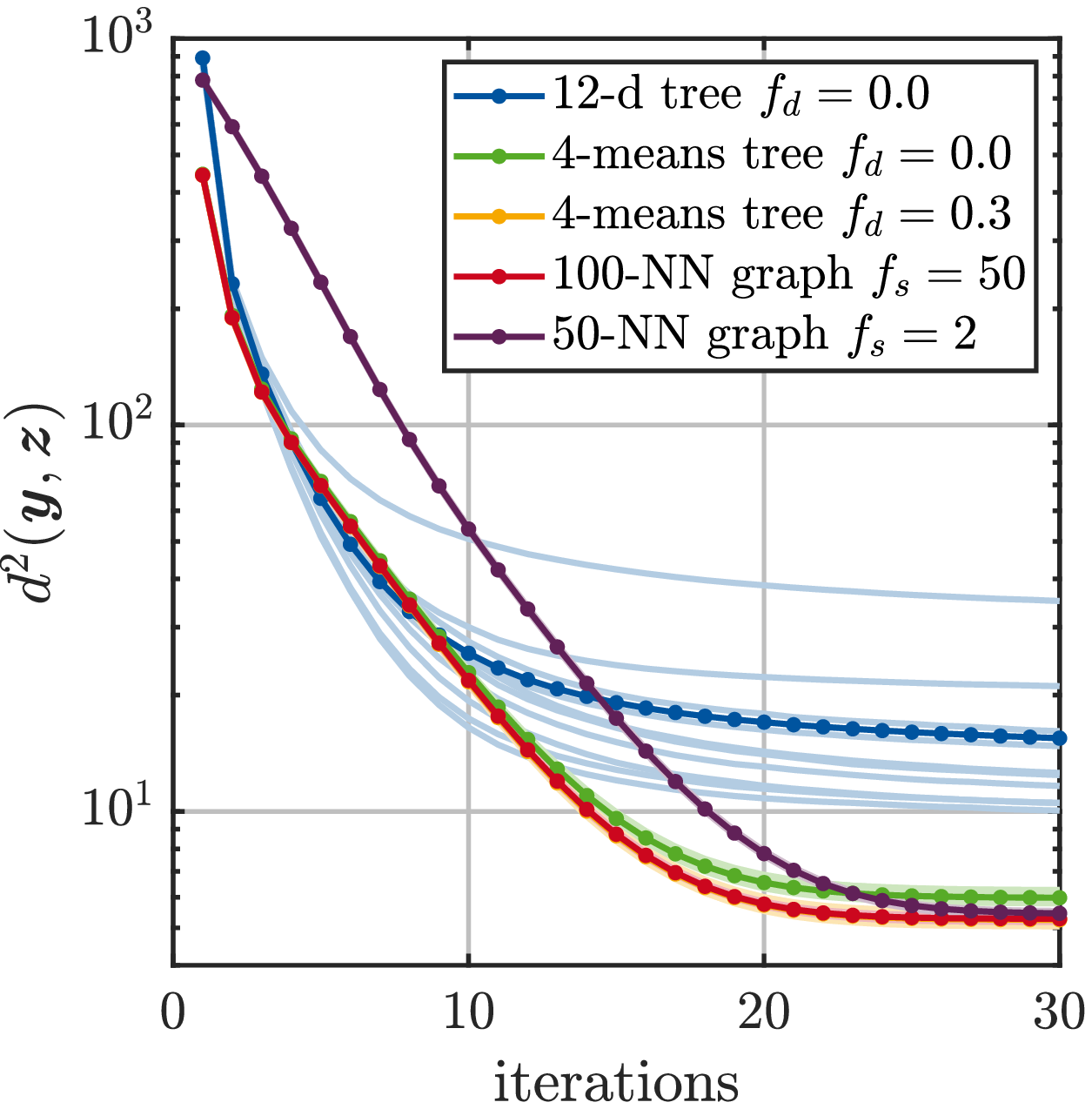}
\subcaption{}
\label{fig:distIter100million}
\end{subfigure}
\caption{Elastic solid example: Comparison of different approximate nearest-neighbor algorithms used for nearest-neighbor search on 10 samples of 100 million points data sets. Averaged results are depicted in bold, all results in light colors. a) Query times per iteration over solver iterations. Total search times within the first 30 iterations stated in brackets. b) Global squared distance over solver iterations. Remaining squared distance after 30 iterations stated in brackets.}
\end{figure}

The computations were repeated with the most promising parameter configuration on 100 million points data sets to compare the investigated search algorithms.
Therefore, ten randomly sampled data sets were investigated.
We considered the $k$-d tree with $f_d=0$ and the $k$-means tree with $f_d=0$ and $f_d=0.3$.
Additionally, we chose the $k$-NN graph with $k=100$ and $f_s=50$ as well as $k=75$ and $f_s=1$.
The query times and squared distances over iterations are depicted in Fig.~\ref{fig:timeIter100million} and Fig.~\ref{fig:distIter100million}.
In Tab.~\ref{tab:100mComp} the total query times of the 30 iterations, as well as the remaining distances after 30 iterations, are summarized.
In addition, the averaged times for building and the memory used for storing the data structure are remarked.
Regarding the $k$-NN graph, the index was built using an approximate $k$-means tree with $f_d=0.2$.
Therefore, the graph structure is only an approximation of a $k$-NN graph as in the section above. \\
The results show that the query times for the $k$-d tree and the $4$-means tree with $f_d=0$ are the smallest with on average about 3 seconds.
On the one hand, we observe that the $k$-d tree again shows rather bad results regarding the distances and accuracy.
On the other hand, the accuracy of the $4$-means tree with $f_d=0$ is very promising.
Regarding the final distances, only a small further improvement from 5.98 to 5.25 can be observed considering the results of the $4$-means tree with $f_d = 0.3$.
The smallest remaining distances are computed with the $4$-means tree with $f_d=0.3$, whereas the required query time is also the largest with 105.7 seconds on average.
One further search structure, which seems to be quite efficient is the 75-NN graph with a maximum of two steps in one iteration.
Compared to the 100-NN graph, the algorithm needs only half the time (13.2 seconds) and has mainly the same final squared distances.
Fundamental differences can also be observed concentrating on the building time and needed memory capacity.
On the one hand, the $k$-d tree uses only 111 seconds to build and 0.75 GB to store the index.
On the other hand, the $k$-NN graph needed on average almost 17 hours for building and 37.3 GB for storing.
\begin{table}[htbp]
\centering
\begin{tabular}{P{2.0 cm}|P{1.5 cm}|P{1.5 cm}|P{1.5 cm}|P{1.5 cm}|P{1.5 cm}}
algorithm & parameters & total search time (s) & remaining sqrt. dist & building time (h) & index file size (GBytes)\\ \hline
$k$-d tree     & $f_d=0.0$           & 2.542  & 15.4 & 0.031 & 0.75\\
$k$-means tree & $k = 4$ $f_d = 0.0$ & 3.245  & 5.98 & 0.888 & 1.5\\
$k$-means tree & $k = 4$ $f_d = 0.3$ & 105.7  & 5.25 & 0.888 & 1.5\\
$k$-NN graph   & $k=100$ $f_s=50$    & 27.19  & 5.28 & 16.68 & 37.3\\
$k$-NN graph   & $k=50$ $f_s=2$      & 13.22  & 5.45 & 16.68 & 37.3
\end{tabular}
\caption{Comparison of different indicators for selected search algorithms. Results are averaged over 10 samples.}
\label{tab:100mComp}
\end{table}
%
%
\subsection{k-means tree on a 1 billion points data set}
So far, we observed that the $k$-means tree shows excellent performance concerning query time and accuracy.
Additionally, regarding building time and memory requirements, the tree also dominated the $k$-NN graph structure.
Therefore, we performed a final study using 5 samples with one billion points.
In contrast to the computations in Chapter \ref{sec:kmeans}, the control parameter $f_d$ is not held constant but linearly increased within the first 20 iterations from zero to a final value of $f_d^f$.
Computations were performed for $f_d^f = 0.0, 0.2, 0.4, 0.6$.
The building time for these 4-means trees was on average 12.0 hours and took a memory of 15.5 GB to store a single index.\\
\begin{figure}[!htbp]
\begin{subfigure}{0.5\textwidth}
\centering
\includegraphics[width=\textwidth]{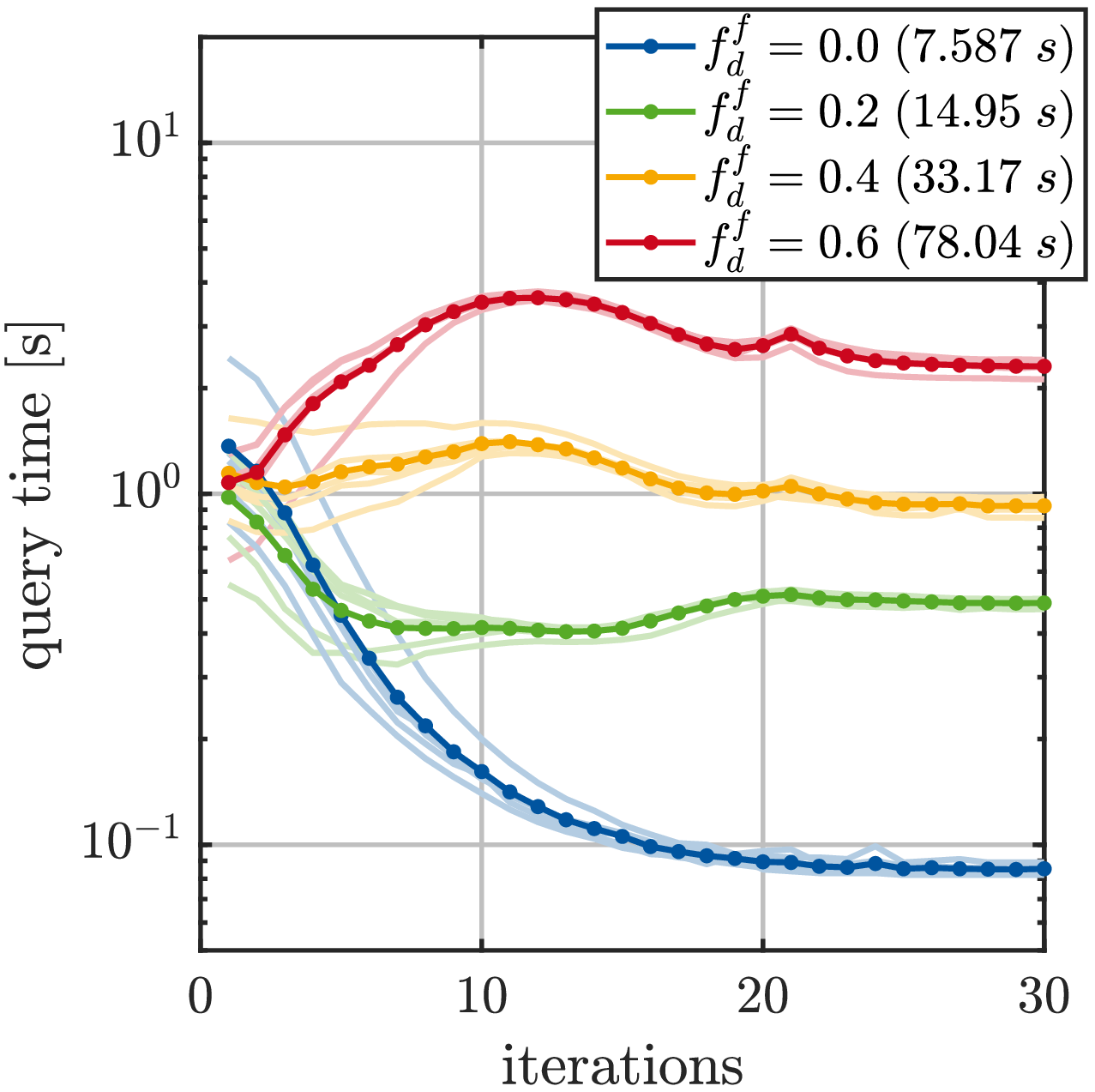}
\subcaption{}
\label{fig:timeIter1billion}
\end{subfigure}
\begin{subfigure}{0.5\textwidth}
\centering
\includegraphics[width=\textwidth]{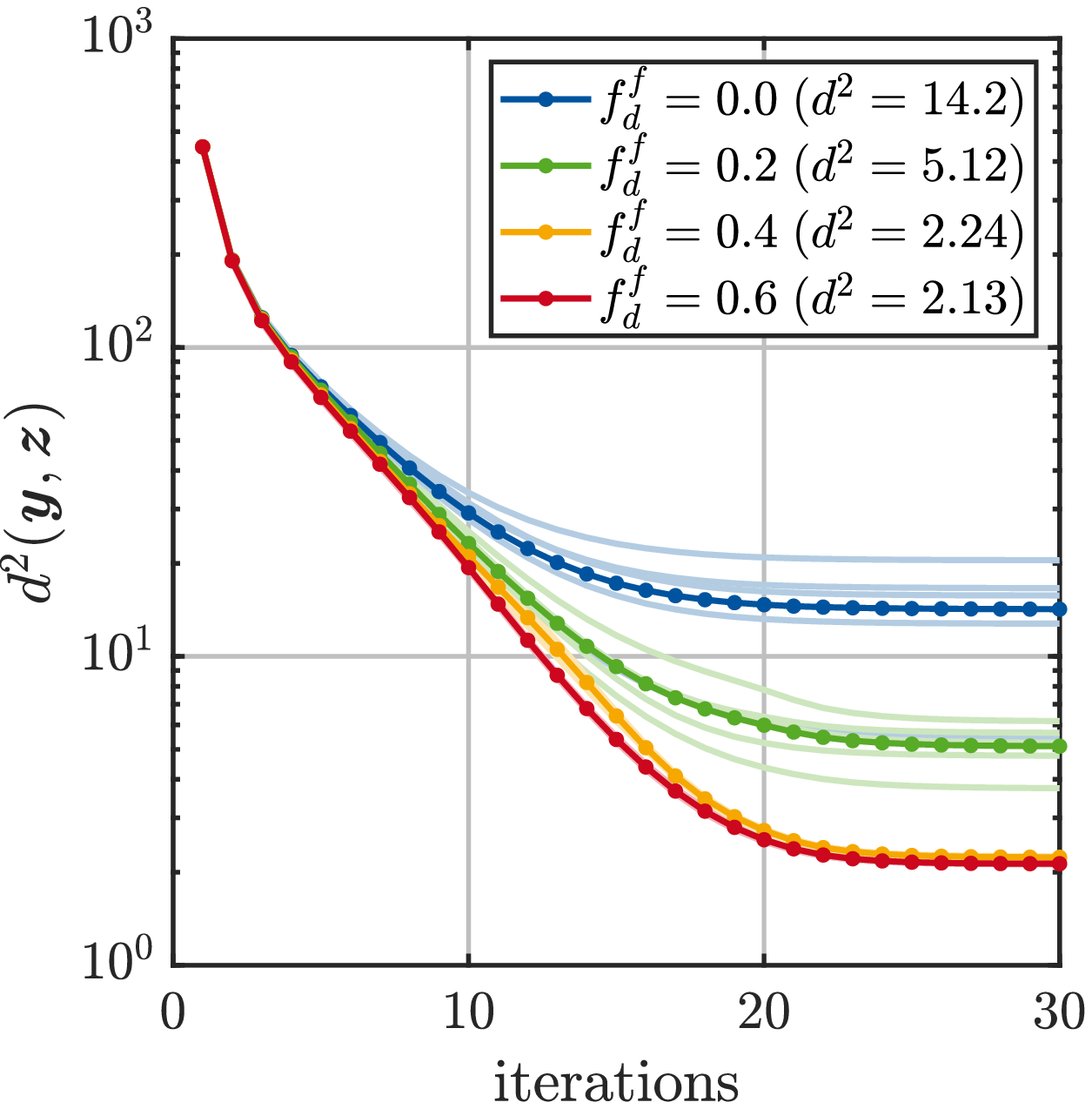}
\subcaption{}
\label{fig:distIter1billion}
\end{subfigure}
\caption{Elastic solid example: Approximate $4$-means tree used for nearest-neighbor search on 5 one billion points data sets. Results are studied for different parameters $f_d^f$. Averaged results are depicted in bold, all results in light colors. a) Query times per iteration over solver iterations. Total search times within the first 30 iterations stated in brackets. b) Global squared distance over solver iterations. Remaining squared distance after 30 iterations stated in brackets.}
\end{figure}
Fig.~\ref{fig:timeIter1billion} and Fig.~\ref{fig:distIter1billion} show the query times and distances over iterations.
In this study, the computations with $f_d^f=0$ took on average 7.6 seconds.
This is already a very impressive result considering that a single computation with the exact $k$-d tree would need approximately 300 days extrapolating from the results obtained in chapter \ref{sec:exactkd}.
That would mean a speedup of more than $10^6$.
Nevertheless, a reduction of accuracy is observed compared to the results with $f_d^f=0.4$ and $f_d^f=0.6$.
The latter show approximately the same remaining distances with $2.24$ and $2.13$.
With a parameter of $f_d^f=0.4$ we end up with approximately 33 seconds per computation.

\section{Summary and concluding remarks} \label{sec:conclusion}

We have investigated three search structures for solving the nearest-neighbor problem in DD computing, namely, the $k$-d tree, the $k$-means tree, and the $k$-NN graph and shown how these search structures can be used for approximate nearest-neighbor searching. 
$k$-means search trees are found to be particularly well-suited for the task because the space partitioning is similar to a Voronoi tessellation. 
We find that, in the initial iterations of the DD solver, rough approximations of the nearest-neighbors are sufficient. 
As convergence is approached, the knowledge of the previous iteration can be used to accelerate the searches, e.~g., using graph structures. 
We show that these strategies allow finite element simulations with data sets of up to billion points to be performed in the span of seconds.

\textit{Further improvements:} With reference to graph search algorithms, we have observed that large benefits can be obtained by taking into account query results from previous iterations. 
The use of that information in the context of $k$-means trees naturally suggests itself. 
We have performed additional assessments of the satellite system graph \cite{Fu:2019} and the hierarchical navigable small world graph \cite{Malkov:2018}. 
For the problems under consideration, we have not observed significant advantages, especially considering the additional effort required for building the indices. 
However, additional performance could be gained by means of parallel implementations or GPU computing.

\textit{Noisy data sets:} We have limited our investigations to 3D elastic case problems with synthetic data sets free of scatter or noise. 
Such data sets might be generated, e.~g., by simulations at a lower scale or by preprocessing or filtering measured data.
However, experimetnal data sets are inevitably noisy and may contain outliers.
In \cite{Kirchdoerfer:2017}, a maximum-entropy solver was proposed that can deal effectively with noisy data and outliers. 
The resulting solvers are no longer minimum-distance solvers by aim to maximize a likelihood function. 
Nevertheless, the need to structure the data efficiently and perform fast searches persists.
Given the maximum entropy characterization of the solutions, $k$-mean tree methods are likely to be best suited.

\textit{Connection to machine learning:} The search data structures discussed in this paper may be regarded as a form of set-oriented unsupervised machine learning. 
Thus, in contrast to other forms of supervised machine learning, such as Artificial Neural Networks, here the outcome of the learning is not a functional relation, e.~g., between stress and strain, but the data structures themselves. 
Such data structures reveal, or allow us to learn, how the data is organized. 
Unlike Artificial Neural Networks, the unsupervised learning afforded by the search data structures is \textit{lossless}, i.~e., it entails no loss of information relative to the original data set. 
In \cite{Lin:2006}, the connection between the $k$-NN problem and random decision forests \cite{Ho:1995} was already noted. 
The random decision forest, also known as random forest regression, is a machine learning technique based on multiple decision trees \cite{Breiman:1984}. 
An application of such methods to DD computing immediately suggests itself. 
In a previous publication \cite{Eggersmann:2020}, the authors noted how other unsupervised machine learning techniques such as tensor voting \cite{Mordohai:2010} can be used to set up additional local data structures.

\textit{Data warehouse:} Material data sets, such as generated by DDMI \cite{Leygue:2018, Leygue:2019, Dalemat:2019} can be exceedingly large and place onerous memory demands. 
This suggests centralization, or warehousing, of large material data sets in specialized servers for public access. 
In this data infrastructure, queries from a finite-element calculation performed on a local computer would be sent to the material data warehouse, which would perform near-neighbor searches and other operations on pre-structured data sets and send the results to the local user. 
Today, this procedure is already standard for many applications like web search or mobile navigation applications.
By contrast, this form of data acquisition, management and exchange remains largely to be explored and developed in engineering applications. 

\section*{Acknowledgments}

MO gratefully acknowledges the support of the Deutsche Forschungsgemeinschaft (DFG) through the Sonderforschungsbereich 1060 ``The mathematics of emergent effects". SR and RE gratefully acknowledge the  financial support of the Deutsche Forschungsgemeinschaft (DFG) through the project RE 1057/40-2 ``Model order reduction in space and parameter dimension — towards damage-based modeling of polymorphic uncertainty in the context of robustness and reliability" within the priority program SPP 1886 ``Polymorphic uncertainty modelling for the numerical design of structures". Finally, all authors acknowledge the financial support of the DFG and French Agence Nationale de la Recherche (ANR) through the project ``Direct Data-Driven Computational Mechanics for Anelastic Material Behaviours" (project numbers: ANR-19-CE46-0012-01, RE 1057/47-1) within the French-German Collaboration for Joint Projects in Natural, Life and Engineering (NLE) Sciences.

\bibliography{Efficient_DD_Bib}

\end{document}